\documentclass[%
showpacs,%preprintnumbers,
nofootinbib,
 amsmath,amssymb,
 aps,
prd,
twocolumn,
floatfix,
]{revtex4-1} % {article} % {revtex4-1} 
\pdfoutput=1 %arXiv says to do this for PDFLaTeX submisssion
\usepackage{graphicx}% Include figure files
\usepackage{dcolumn}% Align table columns on decimal point
\usepackage{bm}% bold math
\usepackage{hyperref}% add hypertext capabilities
\usepackage[]{units}
\usepackage[]{nicefrac}
\usepackage{verbatim}
\usepackage{slashed}
\usepackage{color}

\def\slashchar#1{\setbox0=\hbox{$#1$}
   \dimen0=\wd0 \setbox1=\hbox{/} \dimen1=\wd1
   \ifdim\dimen0>\dimen1 \rlap{\hbox to \dimen0{\hfil/\hfil}} #1
   \else  \rlap{\hbox to \dimen1{\hfil$#1$\hfil}} / \fi}

\usepackage{multirow}
\usepackage{mathtools}
\usepackage{amsfonts} % AMS
\usepackage{amssymb} % AMS
\usepackage{amsmath} % AMS
\usepackage{subfigure} % Include figure files
\usepackage{array} % array
\usepackage{dcolumn} % Align table columns on decimal point
\usepackage{bm} % bold math
\usepackage{latexsym} % latex symbols
\usepackage{longtable} % long tables
\usepackage{epsfig}

\begin{document}

%\preprint{MITP, HIM}

\title{On the effect of excited states in lattice calculations of the nucleon axial charge }
\author{Maxwell T. Hansen$^a$}
 \email{hansen@kph.uni-mainz.de}
\author{Harvey B.\ Meyer$^{a,b}$}%
 \email{meyerh@kph.uni-mainz.de}
\affiliation{$^a$Helmholtz~Institut~Mainz, D-55099 Mainz, Germany \\ 
$^b$PRISMA Cluster of Excellence and Institut f\"ur Kernphysik 
Johannes Gutenberg-Universit\"at Mainz, D-55099 Mainz, Germany }

\date{\today}

\begin{abstract}
Excited-state contamination is one of the dominant uncertainties in lattice calculations of the nucleon axial-charge, $g_A$. Recently published results in leading-order chiral perturbation theory (ChPT) predict the excited-state contamination to be independent of the nucleon interpolator and positive \cite{BrianGA,BarTwoPoint,BarGA}. However, empirical results from numerical lattice calculations show negative contamination (downward curvature), indicating that present-day calculations are not in the regime where the leading-order ChPT predictions apply. In this paper we show that, under plausible assumptions, one can reproduce the behavior of lattice correlators by taking into account final-state $N \pi$ interactions, in particular the effect of the Roper resonance, and by postulating a sign change in the infinite-volume $N \to N \pi$ axial-vector transition amplitude.

% May be entered using the \verb+\pacs{#1}+ command.
%\pacs{11.15.Ha, 12.38.Gc}% PACS, the Physics and Astronomy
\end{abstract}

\maketitle

%%%%%%%%%%%%%%%%%%%%%%%%%%%
%%%%%%%%%%%%%%%%%%%%%%%%%%%
%%%%%%%%%%%%%%%%%%%%%%%%%%%
%%%%%               INTRODUCTION            %%%%%
%%%%%%%%%%%%%%%%%%%%%%%%%%%
%%%%%%%%%%%%%%%%%%%%%%%%%%%
%%%%%%%%%%%%%%%%%%%%%%%%%%%

\section{\label{sec:introduction}Introduction}

In the past decades, outstanding progress has been made in reproducing the properties of the strong interaction by numerically calculating QCD correlators on a Euclidean spacetime lattice. One goal of such calculations is to extract various aspects of nuclear structure from the underlying theory, and a target quantity here is the nucleon axial charge, $g_A$, defined via
\begin{equation}
\label{eq:gAdef}
\langle N, \textbf p,\sigma' \vert A^{a}_\mu(0) \vert N, \textbf p, \sigma \rangle = g_A \overline u_{\sigma'}(\textbf p) \Gamma_{A,\mu}^{a} u_{\sigma}(\textbf p) \,,
\end{equation}
where $A^{ a}_\mu \equiv \overline {\mathcal Q}\Gamma_{A,\mu}^{a } \mathcal Q$ and $\Gamma_{A,\mu}^{a } =  T^a \gamma_\mu \gamma_5$. Here $T^a = \tau^a/2$ are the generators of $SU(2)$ isospin and $\mathcal Q$ is a doublet containing the up and down quarks. We have also introduced single nucleon states with momentum $\textbf p$ and spin $\sigma, \sigma'$ as well as their corresponding spinors $\overline u_{\sigma'}(\textbf p)$ and $ u_{\sigma}(\textbf p)$. These are also isospin doublets built from the proton and neutron. In this work we use Euclidean conventions for the gamma matrices, $\{\gamma_\mu , \gamma_\nu  \}=2 \delta_{\mu \nu}$.

The axial charge is in many ways an ideal quantity for lattice QCD (LQCD). In particular, it can be directly accessed from plateaus in Euclidean correlators and does not contain the noisy quark-disconnected diagrams. However, as a nuclear quantity, it suffers from the signal-to-noise problem and this is only made worse in the three-point function required to create a nucleon, couple it to the axial current and then annihilate it. For some time now, lattice calculations of $g_A$ have been prone to underestimate the quantity.%
%%%%%%%%%
% FOOTNOTE
%%%%%%%%%
\footnote{See, for example, Refs.~\cite{MarthaRev2014,GreenRev2016,ETMAvx2016,BhattacharyaAxial2016,QCDSFAxial2015,EigoAMA2016}.} %
%%%%%%%%%
% FOOTNOTE
%%%%%%%%%
Possible explanations for this include underestimated systematic uncertainties from extrapolation to the physical pion mass, from finite-volume effects and from excited-state contamination. This work is concerned with the latter.

Specifically we are interested in excited-state contamination in the context of a ratio of Euclidean three- and two-point correlators, constructed to satisfy
\begin{multline}
\label{eq:introesc}
R(T,t) \underset{T \gg t \gg 0}{\longrightarrow} g_A \\ + \sum_{n=2} \left ( b_n \big (e^{- \Delta E_n (T - t)}+  e^{- \Delta E_n t} \big ) + c_n e^{-  \Delta E_n T} \right ) \,,
\end{multline}
where we have dropped subleading exponentials as we explain in detail in the following section. Here we introduce $T$ as the nucleon source-sink separation, $t$ as the current insertion time, and $\Delta E_n$ as the gap between the nucleon mass and the $(n-1)$th excited state. The coefficients $b_n$ and $c_n$ are related to finite-volume matrix elements as given in Eqs.~(\ref{eq:bndef}) and (\ref{eq:cndef}) below.  

The excited-state contribution to $R(T,t)$ has been recently studied in both non-relativistic \cite{BrianGA} and relativistic \cite{BarTwoPoint,BarGA} baryon chiral perturbation theory (ChPT). In both cases the authors find that the leading-order (LO) ChPT predictions are independent of the form of the nucleon interpolators.%
%%%%%%%%%
% FOOTNOTE
%%%%%%%%%
\footnote{This assumes local three-quark operators. As is carefully discussed in Ref.~\cite{BarTwoPoint}, the prediction also holds for smeared operators, provided that the smearing radius is sufficiently small.} %
%%%%%%%%%
% FOOTNOTE
%%%%%%%%%
This leads to the universal prediction that $b_n>0$, and thus that the excited-state contamination is positive. Since the predictions for $b_n$ and $c_n$ depend only on $g_A$, the pion decay constant, $f_\pi$, and known kinematic quantities, the ChPT expressions could in principle be used to remove the leading excited-state contribution in order to more accurately extract $g_A$.

To make use of the LO ChPT results, however, one must ensure that these describe present-day numerical LQCD data. As $g_A$ is often extracted from the central value, $R(T,T/2)$, or by fitting a constant to a range of central values, determining the $T$ values needed for $R(T,T/2)$ to enter the regime of LO ChPT is particularly useful. If the source-sink separation is too small, then the set of finite-volume states needed to estimate $R(T,T/2)$ goes beyond the region described by the leading-order prediction. Indeed, the curvature of nearly all available numerical LQCD data for $R(T,T/2)$ as a function of $T$ is negative, indicating negative excited-state contamination, in contradiction with the LO ChPT prediction.%
%%%%%%%%%
% FOOTNOTE
%%%%%%%%%
\footnote{Again see Refs.~\cite{MarthaRev2014,GreenRev2016,ETMAvx2016,BhattacharyaAxial2016,QCDSFAxial2015,EigoAMA2016}. One exception here is the curvature of the correlator data of Ref.~\cite{Ohta2015}. It is unclear why the results of this work differ from the rest. One possibility is that, as compared to other calculations, the interpolators used in this study have enhanced coupling to the lower excited states.} %
%%%%%%%%%
% FOOTNOTE
%%%%%%%%%
Similarly, at fixed $T$, $R(T,t)$ is consistently observed to have negative curvature as a function of the current-insertion time, $t$. We take this as strong evidence that, in present day LQCD calculations, the values of $T$ are too small for $R(T,t)$ to be well described by the LO ChPT results.

In this paper we show that, under plausible assumptions, one can reproduce the qualitative behavior of numerical LQCD correlators by including the contributions of higher-energy states, taking into account $N \pi$ final-state interactions, and postulating a sign change in the infinite-volume axial-vector transition amplitude, $\langle N \pi, \mathrm{out} \vert A_\mu \vert N \rangle$. Using experimentally-determined $N \pi$ scattering data in a generalization of L\"uscher's quantization condition \cite{Luscher1986, Luscher1990}, we predict the energies of the finite-volume excited states entering $\Delta E_n$. We then use a generalization of the Lellouch-L\"uscher formalism, again with experimental scattering data, to relate the finite-volume matrix elements in $b_n$ and $c_n$ to infinite-volume matrix elements involving $N \pi$ asymptotic states \cite{Lellouch2000}. To complete the construction, we estimate the remaining infinite-volume matrix elements in a model based on LO ChPT, supplemented by the scattering data. 

Within this set-up we find that a large number of excited states give an important contribution to $R(T,T/2)$ for realistic $T$ values, and that a sign flip in the axial-vector transition can readily accommodate the empirically observed negative excited-state contamination [see Figs.~\ref{fig:esc1} and \ref{fig:esc2} below]. We find that, for physical pion masses, $T \gtrsim 2 \mathrm{\, fm}$ is needed to enter the regime where LO ChPT describes the lattice correlators.

This analysis suffers from various limitations that prevent us from offering reliable quantitative predictions. The most important limitation is the neglect of $N \pi \pi$ states. Here we only study the energies and matrix elements of finite-volume $N \pi$ states. Both the L\"uscher quantization condition and the Lellouch-L\"uscher formalism hold only for energies below three-particle production threshold, but in this work we also include energies above $N \pi \pi$ threshold where the relations develop uncontrolled systematic uncertainties. There is evidence that the breakdown of the formalism turns on slowly as one crosses multi-particle thresholds,%
%%%%%%%%%
% FOOTNOTE
%%%%%%%%%
\footnote{See, for example, the phase-shifts extracted above multi-particle thresholds in Ref.~\cite{WilsonCoupRho2015}.} %
%%%%%%%%%
% FOOTNOTE
%%%%%%%%%
but in the vicinity of the Roper resonance, the neglected three-particle states could have a significant contribution. [See also the discussion in the paragraph following Eq.~(\ref{eq:omNfree}) below.] Other limitations of this study include the modeling of the infinite-volume matrix elements, explained in detail in Sec.~\ref{sec:infME}, as well as the restriction to physical pion masses. The latter is a natural limitation given our approach of working with experimental scattering data. As a result, the predictions for $R(T,t)$ discussed in Sec.~\ref{sec:contam} are most directly applicable to ensembles near the physical point. 

As an aside we comment that, in order to have a solid theoretical foundation for this work, it was necessary to make contact with the LO ChPT results derived in Refs.~\cite{BarTwoPoint,BarGA}. In these earlier publications, $\Delta E_n$ is approximated using non-interacting $N \pi$ states in a finite volume, so that the work is concerned only with predicting the coefficients, $b_n$ and $c_n$. Since we are using the Lellouch-L\"uscher formalism to predict the coefficients in this study, it was necessary to first understand how this formalism can be used to reproduce the LO ChPT results. We were able to make this connection in detail, re-deriving some of the expressions reported in Refs.~\cite{BarTwoPoint,BarGA}. This is interesting in its own right as it shows how the Lellouch-L\"uscher formalism provides a shortcut for extracting ChPT predictions of these and related quantities. In particular, the numerous one-loop diagrams needed to determine $b_n$ in Ref.~\cite{BarGA} are replaced in the present approach by five tree-level diagrams. Details are given in the appendix.

The remainder of this article is organized as follows. In the following section we define the correlators, the ratio $R$ and the parameters $\Delta E_n$, $b_n$ and $c_n$ that describe the excited states. In Sec.~\ref{sec:es} we use experiemental partial wave data to estimate the interacting energy gaps $\Delta E_n$ associated with $N \pi$ states. Then in Sec.~\ref{sec:LL} we give estimates for the coefficients $b_n$ and $c_n$. This leads to estimates of the excited state contamination for typical present-day lattice set-ups, presented in Sec.~\ref{sec:contam}. In the appendix we detail the derivation of various ChPT expressions used in the main text.

%%%%%%%%%%%%%%%%%%%%%%%%%%%
%%%%%%%%%%%%%%%%%%%%%%%%%%%
%%%%%%%%%%%%%%%%%%%%%%%%%%%
%%%%%                       RATIO                     %%%%%
%%%%%%%%%%%%%%%%%%%%%%%%%%%
%%%%%%%%%%%%%%%%%%%%%%%%%%%
%%%%%%%%%%%%%%%%%%%%%%%%%%%

\section{\label{sec:extractGA}Extracting $g_A$ from the lattice}

Various methods exist for using numerical LQCD to determine $g_A$. Common to all approaches is the determination of two- and three-point correlators of the form
\begin{align}
\begin{split}
\label{eq:C3def}
C^{}_3(T,t) & \equiv \int d^3 \textbf x \int d^3 \textbf y \   \Gamma'_{\mu, \alpha \beta}  \\
& \hspace{50pt} \times \langle \mathcal O_\beta(\textbf x, T) A^{3}_\mu(\textbf y, t) \overline {\mathcal O}_\alpha(0) \rangle \,,  \end{split} \\
C^{}_2(T) & \equiv  \int d^3 \textbf x \   \Gamma_{\alpha \beta} \langle \mathcal O_\beta(\textbf x, T)   \overline {\mathcal O}_\alpha(0) \rangle \,,
\label{eq:C2def}
\end{align}
where $\overline {\mathcal O}_\alpha$, ${\mathcal O}_\beta$ are proton interpolating fields, $A^{3}_\mu$ is the third isospin component of the axial vector current, and $\Gamma'$ and $\Gamma$ are projectors. In this work we restrict attention to states that have zero three-momentum in the finite-volume frame.

Defining $\widetilde {\mathcal O}^{}_\beta(T) \equiv  \int d^3 \textbf x \  \mathcal O_\beta(\textbf x, T)$,\\ $\widetilde A^{3}_\mu(t) \equiv  \int d^3 \textbf y \ A^{3}_\mu(\textbf y, t)  $, and performing a spectral decomposition, we reach
\begin{align}
\label{eq:sd3}
\begin{split}
C^{}_3(T,t) & \equiv L^{-3}  \sum_{n,m} \ \Gamma'_{\mu, \alpha \beta}  \langle 0 \vert \widetilde {\mathcal O}^{}_\beta \vert n  \rangle \langle n  \vert \widetilde A^{3}_\mu \vert m \rangle \\[-10pt] &  \hspace{60pt}  \times \langle m \vert  \widetilde {\overline {\mathcal O}}^{}_\alpha \vert 0 \rangle  e^{- E_n(T-t)} e^{- E_m t} \,,
\end{split} \\[5pt]
C^{}_2(T) & \equiv  L^{-3} \sum_{n} \  \Gamma_{\alpha \beta} \langle 0 \vert \widetilde {\mathcal O}^{}_\beta \vert n \rangle  \langle n \vert  \widetilde{ \overline {\mathcal O}}^{}_\alpha \vert 0 \rangle e^{- E_n T} \,,
\label{eq:sd2}
\end{align}
where we have assumed $T>t>0$ and have used the shorthand $\widetilde {\mathcal O}^{}_\beta \equiv \widetilde {\mathcal O}^{}_\beta(0)$ and similar for $\widetilde A^{3}_\mu$. To treat the fields equivalently we have Fourier transformed $\overline {\mathcal O}_\alpha$ over spatial volume but have also divided by volume to preserve the definitions. Throughout this work all finite-volume states are normalized as $\langle n \vert n \rangle = 1$.

We next observe that the lowest state in the sum, denoted by $n,m=1$, is the single nucleon state. From this follows that the ratio of the $n,m=1$ terms in $C_3(T,t)$ and $C_2(T)$ gives $g_A$
\begin{equation}
g_A \equiv \frac{\Gamma'_{\mu, \alpha \beta}  \langle 0 \vert \widetilde {\mathcal O}^{}_\beta \vert 1 \rangle \langle 1 \vert \widetilde A^{3}_\mu \vert 1 \rangle \langle 1 \vert  \widetilde {\overline {\mathcal O}}^{}_\alpha \vert 0 \rangle}{\Gamma_{\alpha \beta} \langle 0 \vert  \widetilde {\mathcal O}^{}_\beta \vert 1 \rangle \langle 1 \vert   \widetilde{\overline {\mathcal O}}^{}_\alpha \vert 0 \rangle} \,.
\end{equation}
This relies on the definitions of $\Gamma$ and $\Gamma'$. These are constructed to ensure that the result holds.

It follows that $g_A$ can be accessed by identifying a plateau in the ratio
\begin{equation}
R^{}(T,t) \equiv \frac{C^{}_3(T,t)}{C^{}_2(T)} \,.
\end{equation}
Substituting the spectral decompositions, Eqs.~(\ref{eq:sd3}) and (\ref{eq:sd2}), taking $T \gg t \gg 0$ and expanding the denominator, we find
\begin{multline}
\label{eq:Rdecom}
R^{}(T,t) = g_A + \sum_{n=2}^\infty \bigg [ b_n \big ( e^{- \Delta E_n (T - t)} +  e^{- \Delta E_n t}  \big )  \\   + c_n e^{- \Delta E_n T } + \cdots   \bigg ] \,,
\end{multline}
where $\Delta E_n \equiv E_n - E_1 = E_n -  m_N  + \mathcal O(e^{- M_\pi L})$, with $E_n$ the energy of the $(n-1)$th excited state, $m_N$ the nucleon mass and $M_\pi$ the pion mass. Here we have introduced $L$ as the linear spatial extent of the volume and have used the fact that finite-volume corrections to the nucleon mass are exponentially suppressed. We neglect such corrections throughout. 

In Eq.~(\ref{eq:Rdecom}) we have also introduced
\begin{align}
\label{eq:bndef}
b_n & \equiv \frac{ \Gamma'_{\mu, \alpha \beta}  \langle 0 \vert {\widetilde {\mathcal O}}^{}_\beta \vert n  \rangle \langle n \vert \widetilde A^{3}_\mu \vert 1 \rangle  \langle 1  \vert \widetilde {\overline {\mathcal O}}^{}_\alpha \vert 0 \rangle}{ \Gamma_{ \alpha \beta}  \langle 0 \vert {\widetilde {\mathcal O}}^{}_\beta \vert 1 \rangle  \langle 1 \vert \widetilde {\overline {\mathcal O}}^{}_\alpha \vert 0 \rangle }    \,,  \\
c_n & \equiv   - g_A c_{2,n} + c_{3,n}  \,, 
\label{eq:cndef}
\end{align} 
where
\begin{align}
c_{2,n} & = \frac{\Gamma_{ \alpha \beta}  \langle 0 \vert {\widetilde {\mathcal O}}^{}_\beta \vert n \rangle \langle n \vert \widetilde {\overline {\mathcal O}}^{}_\alpha \vert 0 \rangle }{\Gamma_{ \alpha \beta}  \langle 0 \vert {\widetilde {\mathcal O}}^{}_\beta \vert 1 \rangle \langle 1 \vert \widetilde {\overline {\mathcal O}}^{}_\alpha \vert 0 \rangle}     \,,
\label{eq:c2ndef}\\
\label{eq:c3ndef}
c_{3,n} & = \frac{\Gamma'_{\mu, \alpha \beta}  \langle 0 \vert {\widetilde {\mathcal O}}^{}_\beta \vert n \rangle \langle n \vert \widetilde A^{3}_\mu \vert n \rangle \langle n \vert \widetilde {\overline {\mathcal O}}^{}_\alpha \vert 0 \rangle}{\Gamma_{ \alpha \beta}  \langle 0 \vert {\widetilde {\mathcal O}}^{}_\beta \vert 1 \rangle \langle 1 \vert \widetilde {\overline {\mathcal O}}^{}_\alpha \vert 0 \rangle} \,.
\end{align}
Note that the definition for $b_n$, Eq.~(\ref{eq:bndef}), directly arises from the coefficient on the first exponential, $ e^{- \Delta E_n (T - t)} $, whereas the factor multiplying the second exponential, $ e^{- \Delta E_n t}$, has a different definition. However, as long as $\Gamma'_{\mu}$ is anti-hermitian and $\Gamma$ is hermitian, then Euclidean definitions of charge-conjugation and time-reversal invariance imply $R(T,t) = R(T,T-t)=R^*(T,t)$. Thus the two coefficients are identically equal and we take $b_n$ as the coefficient for both source-to-current and current-to-sink time dependence.

As can be seen by comparing the definitions, Eqs.~(\ref{eq:bndef}) and (\ref{eq:cndef}), the matrix elements required to access the source-to-sink coefficient, $c_n$, are more complicated than those needed for $b_n$. The first term in the definition of $c_n$, proportional to $c_{2,n}$ defined in Eq.~(\ref{eq:c2ndef}), arises from expanding the excited-state contamination of $C_2(T)$ in the denominator. This term depends on the same matrix elements that appear in the definition of $b_n$ and can be studied using the same approach. The second term in $c_n$, $c_{3,n}$ defined in Eq.~(\ref{eq:c3ndef}), arises from source-to-sink contributions in $C_3(T,t)$ and is thus more complicated. This term turns out to be numerically suppressed in LO ChPT and thus unimportant in our qualitative study. With this in mind, in this work we simply use the LO ChPT result for $c_{3,n}$ and only apply the Lellouch-L\"uscher like analysis to $b_n$ and $c_{2,n}$.

The ellipsis in Eq.~(\ref{eq:Rdecom}) stands for terms suppressed by additional factors of $e^{- \Delta E_m t}$, $e^{- \Delta E_m (T-t)}$ or $e^{- \Delta E_m T}$. These neglected terms arise for two reasons. One contribution is from higher orders in the expansion of $C_2(T)$ in the denominator. This expansion is not required but is a good approximation and simplifies the resulting expressions. The second neglected contribution is from terms in Eq.~(\ref{eq:sd3}) with $n \neq m$, with both indices corresponding to excited states. Such terms involve two-to-two (rather than one-to-two) axial-vector matrix elements and are expected to be suppressed relative to those we keep. We caution that these two-to-two transitions are not necessarily volume suppressed. For example, LO ChPT predicts the same volume dependence for $c_{2,n}$ and $c_{3,n}$. This is the case because, at leading order, the current mediating the two-to-two transition couples only to one of the two particles. When the current couples to both particles an extra factor of volume suppression does arise.%
%%%%%%%%%
% FOOTNOTE
%%%%%%%%%
\footnote{For full details on the generalization of the Lellouch-L\"uscher approach to two-to-two matrix elements with one-to-one subprocesses, see Ref.~\cite{BH2to2}.}%
%%%%%%%%%
% FOOTNOTE
%%%%%%%%%

The aim of this work is to estimate the value of the sum in Eq.~(\ref{eq:Rdecom}) for given $T$ and $t$. In the following section we study $\Delta E_n$ and in Sec.~\ref{sec:LL} we turn to $b_n$ and $c_n$. 

%%%%%%%%%%%%%%%%%%%%%%%%%%%
%%%%%%%%%%%%%%%%%%%%%%%%%%%
%%%%%%%%%%%%%%%%%%%%%%%%%%%
%%%%%                   ENERGIES                 %%%%%
%%%%%%%%%%%%%%%%%%%%%%%%%%%
%%%%%%%%%%%%%%%%%%%%%%%%%%%
%%%%%%%%%%%%%%%%%%%%%%%%%%%

\section{\label{sec:es} Estimating the excited-state energies}

The finite-volume quantization condition derived by L\"uscher~\cite{Luscher1986,Luscher1990} has since been extended to include moving frames, non-identical and non-degenerate particles, coupled two-particle channels, and particles with spin~\cite{Rummukainen1995, KSS2005, Christ2005, Lage2009, Bernard2010, Doring2011, HSmultiLL, BricenoTwoPart2012, Fu2012, Gockeler2012, BricenoSpin, BHOneToTwoSpin}. These extensions can be used to estimate the finite-volume energies that appear in $R(T,t)$. In particular, in the range $m_N + M_\pi  < E_n < m_N + 2 M_\pi$, the finite-volume energies can be determined using the L\"uscher quantization condition by inputting the experimentally determined phase shift for $N \pi$ scattering.

It is useful to consider these energies relative to the energies of non-interacting particles in a finite-volume. The non-interacting levels are determined by constraining the momentum to satisfy $\textbf p = 2 \pi \textbf n/L$, where $L$ is the linear extent of the volume and $\textbf n$ is a three-vector of integers. This constraint is appropriate to a cubic finite spatial volume with periodic boundary conditions, and in this work we restrict ourselves to this simplest set-up.

\begin{figure}
\begin{center}
%75
\vspace{20pt}
\includegraphics[scale=0.45]{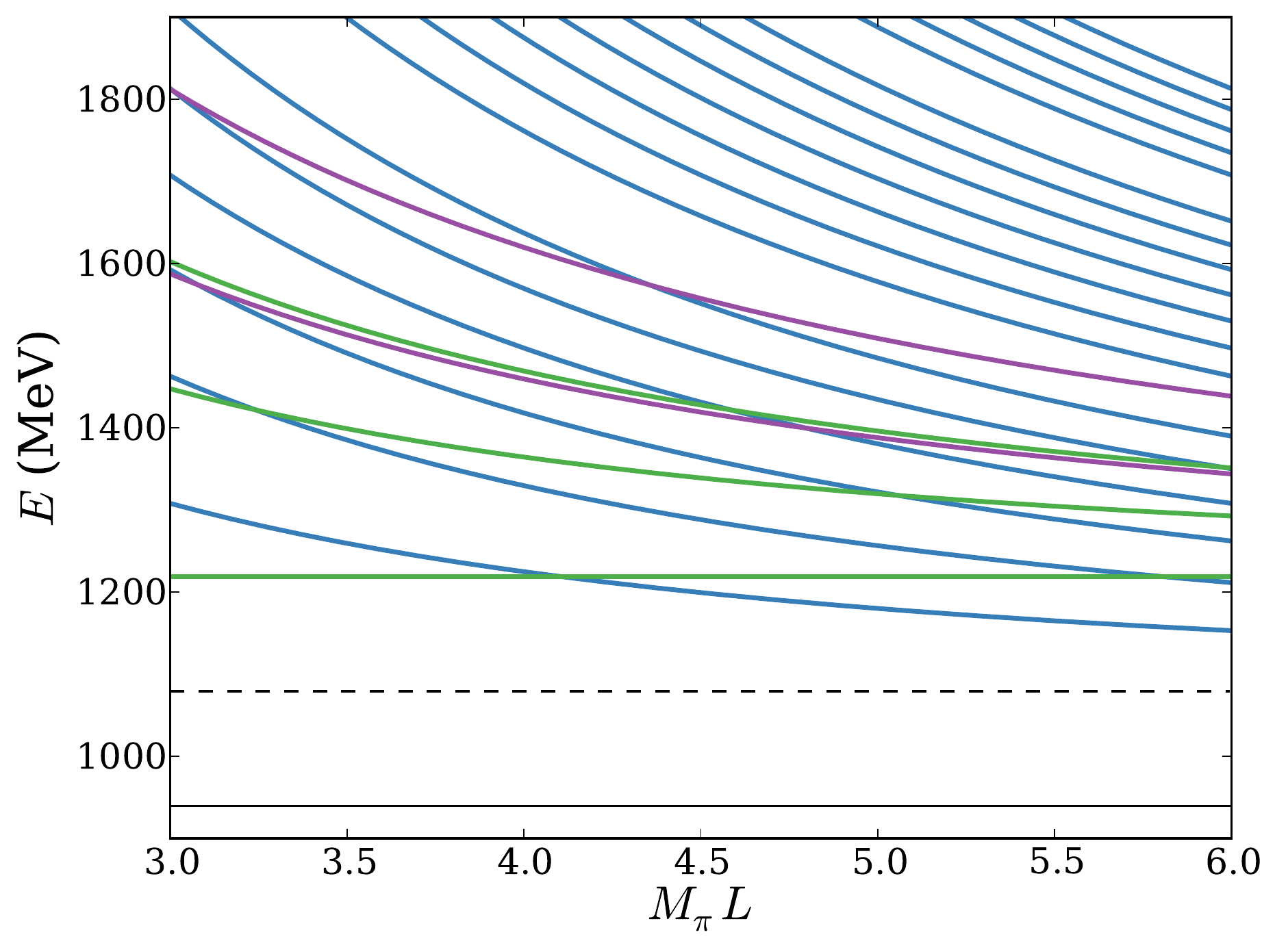}
\end{center}
\caption{Energy levels of non-interacting finite-volume states, with quantum numbers of a single nucleon at rest in the finite-volume frame. The location of the $N \pi$ threshold is indicated by the dashed horizontal line. The state with this energy is not included because its parity is opposite that of the nucleon. The lowest solid horizontal line indicates the single nucleon energy, and the gap from here determines the size of the contributions to $R(T,t)$. Finally, we have included three different types of finite-volume states, distinguished by three colors. Blue levels are back-to-back $N \pi$ states, green levels are $N \pi \pi$ states with one pion at rest, and magenta are $N \pi \pi$ states with the nucleon at rest. For the latter two sets only the first few levels are shown to avoid clutter.}
\label{fig:freelevels}
\end{figure}

In Fig.~\ref{fig:freelevels} we display the non-interacting energies as a function of $M_\pi L$, given by
\begin{multline}
E_n \in \Big \{ \{ \omega_{\pi, \textbf n}  + \omega_{N, \textbf n} \} ,    
 \{  \omega_{\pi, \textbf n}+ \omega_{\pi, \textbf m}  + \omega_{N, \textbf n + \textbf m}  \}, \cdots \Big \} \,,
\end{multline}
where
\begin{align}
\omega_{\pi, \textbf n} & \equiv \sqrt{M_\pi^2 + (2 \pi/L)^2 \textbf n^2} \,, \\[5pt]
  \omega_{N, \textbf n} & \equiv \sqrt{m_N^2 + (2 \pi/L)^2 \textbf n^2} \,, 
  \label{eq:omNfree}
\end{align}
and where the ellipsis indicates four- (or more) particle states. As described in the figure caption, we are interested in states that have the quantum numbers of a single nucleon, $I(J^{P}) = \nicefrac[]{1}{2} (\nicefrac[]{1}{2}^{+})$. For this reason the state with a pion and nucleon both at rest does not contribute. This state only couples to the $s$-wave and thus has negative parity due to the intrinsic parity of the pion.

Also apparent from Fig.~\ref{fig:freelevels} is that, for physical pion masses and realistic values of $M_\pi L$, the L\"uscher formalism only rigorously describes, at most, the first excited state. For $E_n> m_N + 2 M_\pi$, an extended three-particle formalism is required. This has recently been developed by one of us for three-pion states, and the extension to general three-particle systems is underway \cite{LtoK,KtoM}. 

Because the three-particle formalism is not yet directly applicable to $N \pi \to N \pi \pi$, in this work we restrict attention to the two-particle formalism, but also apply it above threshold where the predictions suffer from systematic uncertainties. As we explain more in Sec.~\ref{sec:contam}, an important conclusion of our analysis is that finite-volume states in the vicinity of the Roper resonance can contribute significantly to $R(T,t)$. Given that the Roper has a $\sim\!40\%$ branching fraction to $N \pi \pi$ \cite{PDG}, three-particle states certainly need to be included to offer reliable quantitative predictions in this region. However, barring delicate cancellations between two- and three-particle states, the {qualitative} conclusions presented here are expected to hold. Three-particle contributions may also be enhanced when the energy of an $N \pi$ pair within $N \pi \pi$ is close to the delta resonance. Indeed, in the ChPT analysis of Ref.~\cite{BrianGA} the delta resonance was also included and was found to reduce the value of $R(T,T/2)$. Finally we stress that one can only use the L\"uscher quantization condition with scattering amplitudes that take the unitary form of elastic two-particle scattering, Eq.~(\ref{eq:MJdef}). Thus, in this approximation we must also neglect the inelasticity in the two-particle scattering amplitude. [See also Footnote 6.]   

In the following subsections we present our prediction for the finite-volume energy gaps $\Delta E_n$. First, in Sec.~\ref{sec:qc}, we give the quantization condition for general two-particle systems and show how it can be reduced to describe the $N \pi$ states of interest. Then, in Sec.~\ref{sec:expt}, we use the experimental phase-shift data to predict the finite-volume spectrum.

\subsection{Reducing the quantization condition}

\label{sec:qc}

The quantization condition for particles with spin is a straightforward generalization of L\"uscher's original result. Indeed a wide class of generalizations are all described by the same basic form~\cite{Luscher1986, Luscher1990, Rummukainen1995, KSS2005, Christ2005, Bernard2008, Lage2009, Bernard2010, Doring2011, HSmultiLL, BricenoTwoPart2012, Fu2012, Gockeler2012, BricenoSpin, BHOneToTwoSpin}
\begin{equation}
\label{eq:qc}
\det \Big[\mathcal M^{-1}(E_n) + F(E_n, L)  \Big ] = 0 \,.
\end{equation} 
Here $\mathcal M$ is the two-to-two scattering amplitude and $F$ is a known geometric function. This result describes any two-particle system, with any number of two-particle channels with identical or non-identical particles, degenerate or non-degenerate masses and with arbitrary spin. To describe a specific system one need only specify the exact definitions, and in particular the index space, for $\mathcal M$ and $F$.

As a preliminary example we consider a system with one channel of two non-identical {\em scalars} with masses $M_\pi$ and $m_N$. In this case both $\mathcal M$ and $F$ have two sets of spherical harmonic indices. The scattering amplitude is a diagonal matrix in this space, whose entries are related in the standard way to scattering phase shifts. $F$, by contrast, has on- and off-diagonal entries. This encodes the mixing of partial waves due to the reduced symmetry of the box. $F$ can be written as a sum-integral-difference~\cite{Luscher1986, Luscher1990, Rummukainen1995, KSS2005, Christ2005}  
\begin{multline}
\label{eq:Fmatrix}
F_{\ell', m'; \ell, m}(E, L) \equiv  \bigg[ \frac{1}{L^3} \sum_{\textbf k} - \int \frac{d^3 \textbf k}{(2 \pi)^3} \bigg ]  \\ \times \frac{4 \pi Y^*_{\ell', m'}(\hat{\textbf k} ) Y_{\ell,m}(\hat{\textbf k} ) }{2 \omega_\pi 2 \omega_N (E - \omega_\pi - \omega_N + i \epsilon) } \left ( \frac{k}{p} \right )^{\ell + \ell'} \,,
\end{multline}
where
\begin{align}
\omega_\pi \equiv \sqrt{M_\pi^2 +  k^2} \,, \ \ \omega_N \equiv \sqrt{m_N^2 +   k^2}   \,,
\end{align}
$k = \vert \textbf k \vert$, $\hat {\textbf k} = \textbf k/k$, and the sum runs over all ${\textbf k} = (2\pi/L) {\textbf n}$, ${\textbf n} \in \mathbb{Z}^3$. In Eq.~(\ref{eq:Fmatrix}) an ultraviolet regulator is needed to make the quantity well defined. Since the sum and integral have the same ultraviolet divergence, a universal result is recovered as the regulator is removed. Here $p$ is the magnitude of CM frame momentum for particles with energy $E$ and masses $M_\pi$ and $m_N$
\begin{equation}
\label{eq:pdef}
 E \equiv \sqrt{M_\pi^2 + p^2} + \sqrt{m_N^2 + p^2}  \,.
\end{equation}

To incorporate spin in this system it is most straightforward to first work in the basis where the nucleon is polarized along some fixed direction in its CM frame. This new degree of freedom, denoted by $\sigma$, can be accommodated with two simple modifications. First, the amplitude gains an additional index, $\mathcal M = \mathcal M_{\ell', m', \sigma'; \ell, m, \sigma}$. Second, the kinematic matrix $F$ is multiplied with a Kronecker delta, $\delta_{\sigma' \sigma}$. This completely defines the scalar-nucleon quantization condition. Indeed, the arbitrary-spin quantization condition is given by simply multiplying the $F$ matrices with Kronecker deltas \cite{BricenoSpin,BHOneToTwoSpin}.

\bigskip

Next, to connect with experimental phase shifts, it is convenient to change to the basis of total angular momentum, $J$, orbital angular momentum, $\ell$, and azimuthal component of total angular momentum, $\mu$. The basis change is effected by contracting both sets of indices with standard Clebsch-Gordan coefficients. The amplitude in the new basis can be written%
%%%%%%%%%
% FOOTNOTE
%%%%%%%%%
\footnote{Above three-particle threshold this expression no longer applies and an additional parameter must be introduced to parametrize the inelasticity. Here we are neglecting the inelasticity, even above multi-particle threshold. This approximation, which is consistent with the neglect of $N \pi \pi$ states in the L\"uscher quantization condition, breaks down as the energy increases.
}
%%%%%%%%%
% FOOTNOTE
%%%%%%%%%
\begin{equation}
\label{eq:MJdef}
\mathcal M_{J', \ell', \mu'; J, \ell, \mu} \equiv \delta_{J' J} \delta_{\ell' \ell} \delta_{\mu' \mu} \frac{8 \pi E}{p  \cot \delta_{J,\ell}(p) - i p} \,.
\end{equation}
Note that the conservation of orbital angular momentum is special to the meson-baryon system. Generally orbital angular momenta will mix, but in this case conservation of total angular momentum implies that $\ell$ could at most couple with $\ell \pm 1$. Since changing by one unit flips parity, this coupling vanishes and $\ell$ is conserved.

$F$ in the new basis is given by~\cite{Gockeler2012,BricenoSpin}  
\begin{multline}
\label{eq:FJdef}
F_{J', \ell', \mu'; J, \ell, \mu} \equiv \\ \sum_{m,\sigma,m'} \langle \ell \ m, \nicefrac12 \ \sigma \vert J \mu \rangle \langle \ell' \ m', \nicefrac12 \ \sigma \vert J' \mu' \rangle   F_{  \ell', m';   \ell, m} \,.
\end{multline}

We make one final simplification before introducing approximations. One can show that the imaginary parts of $\mathcal M^{-1}$ and $F$ perfectly cancel in Eq.~(\ref{eq:qc}), giving
\begin{equation}
\det \Big [ \overline F_{J' \ell' \mu'; J \ell \mu} +  \delta_{J' J} \delta_{\ell' \ell} \delta_{\mu' \mu} \cot \delta_{J,\ell}(p) \Big ] = 0 \,,
\end{equation}
where $\overline F = 8 \pi E \mathrm{Re}[F] /p$.

We now reduce the quantization condition to a determinant of a finite-dimensional matrix by ignoring high partial waves. It turns out that, in the even-parity sector, we reach the simplest possible truncation by neglecting $\delta_{J,\ell}$ for $\ell \geq 3$. Then the system is truncated to the $\ell=1$ space. In this space $\overline F_{J' \ell' \mu'; J \ell \mu}$ is a $6 \times 6$ matrix: a $2 \times 2$ block for $\ell=1,J=1/2$, a $4 \times 4$ block for $\ell=1,J=3/2$. To determine its specific form we first note that 
\begin{equation}
\overline F_{\ell'=1, m'; \ell=1, m} = - \frac{1}{q \pi^{3/2}}  Z_{00}(1,q^2) \delta_{m'm}  \,,
\end{equation}
where $Z_{00}$ is the L\"uscher zeta-function described in Ref.~\cite{Luscher1990} and $q \equiv p L/(2 \pi)$. The fact that $\overline F$ is proportional to the identity matrix in the $\ell=1$ subspace is preserved when we change to the $J$ basis. Thus, both matrices in the quantization condition are diagonal and the final result is two independent, one dimensional equations
\begin{equation}
\overline F_{11;11} + \cot \delta_{J, \ell=1}(p) = 0 \,,
\end{equation}
for $J=1/2$ or $3/2$. These can be reexpressed as
\begin{equation}
\label{eq:simplestqc}
\phi(q) + \delta_{J, \ell=1}(p)  = n \pi \,,
\end{equation}
where $n$ is an arbitrary integer and
\begin{equation}
\label{eq:phidef}
\cot \phi(q) = \overline F_{11;11} = - \frac{1}{q \pi^{3/2}}  Z_{00}(1,q^2) \,.
\end{equation}
We comment that this has the same form as the s-wave, scalar quantization condition. The quantity $\phi$ is often referred to as the pseudophase.

The fact that the $J=1/2$ and $J=3/2$ sectors decouple can be explained by examining the symmetry group of the finite-volume system. For the case of one scalar and one spin-half particle in a finite cubic box with zero total momentum, the symmetry group is ${}^2O \otimes S_2$ and the irreps are $G_1^{\pm}, G_2^{\pm}$ and $H^{\pm}$~\cite{Bernard2008}. If we neglect $\ell \geq  3$ and thus also neglect $J \geq 5/2$, then we find a perfect correspondence between finite- and infinite-volume irreps $G_1^- \equiv (J=1/2)$ and $H^- \equiv (J=3/2)$. This implies that, within this approximation, the two partial-waves cannot mix, as we have seen by explicit calculation.

\subsection{Predicting the spectrum from the experimental $N \pi$ phase shift}

\label{sec:expt}

\begin{figure}
\includegraphics[scale=0.45]{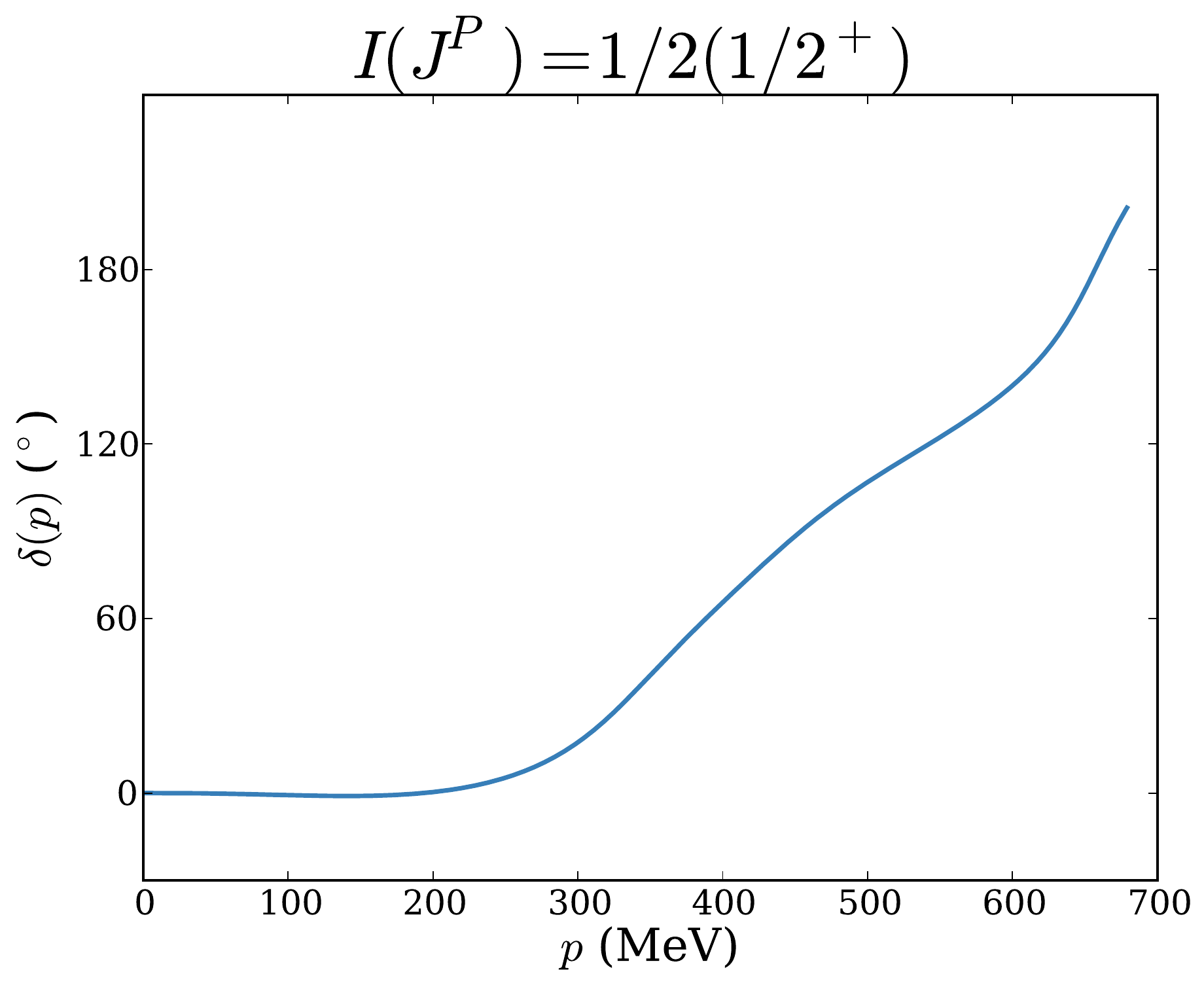}
\caption{The experimental phase shift for $N \pi$ scattering with $I(J^{P}) = \nicefrac[]{1}{2} (\nicefrac[]{1}{2}^{+})$. The slow rise through $\pi/2$ is associated with the broad Roper resonance.}
\label{fig:pshifts}
\end{figure}

To predict the finite-volume spectrum of $N \pi$ states we use experimental data made available by the George Washington University Institute for Nuclear Studies. Their data analysis center is available online at \url{http://gwdac.phys.gwu.edu/analysis/pin_analysis.html}. In this study we use their partial wave analysis WI08 solution. The relevant phase shift data are plotted in Fig.~\ref{fig:pshifts}. For detailed information about the experimental data set and the WI08 fit solution see Refs.~\cite{arndt_extended_2006, paris_toward_2010}.

\begin{figure}
\includegraphics[scale=0.45]{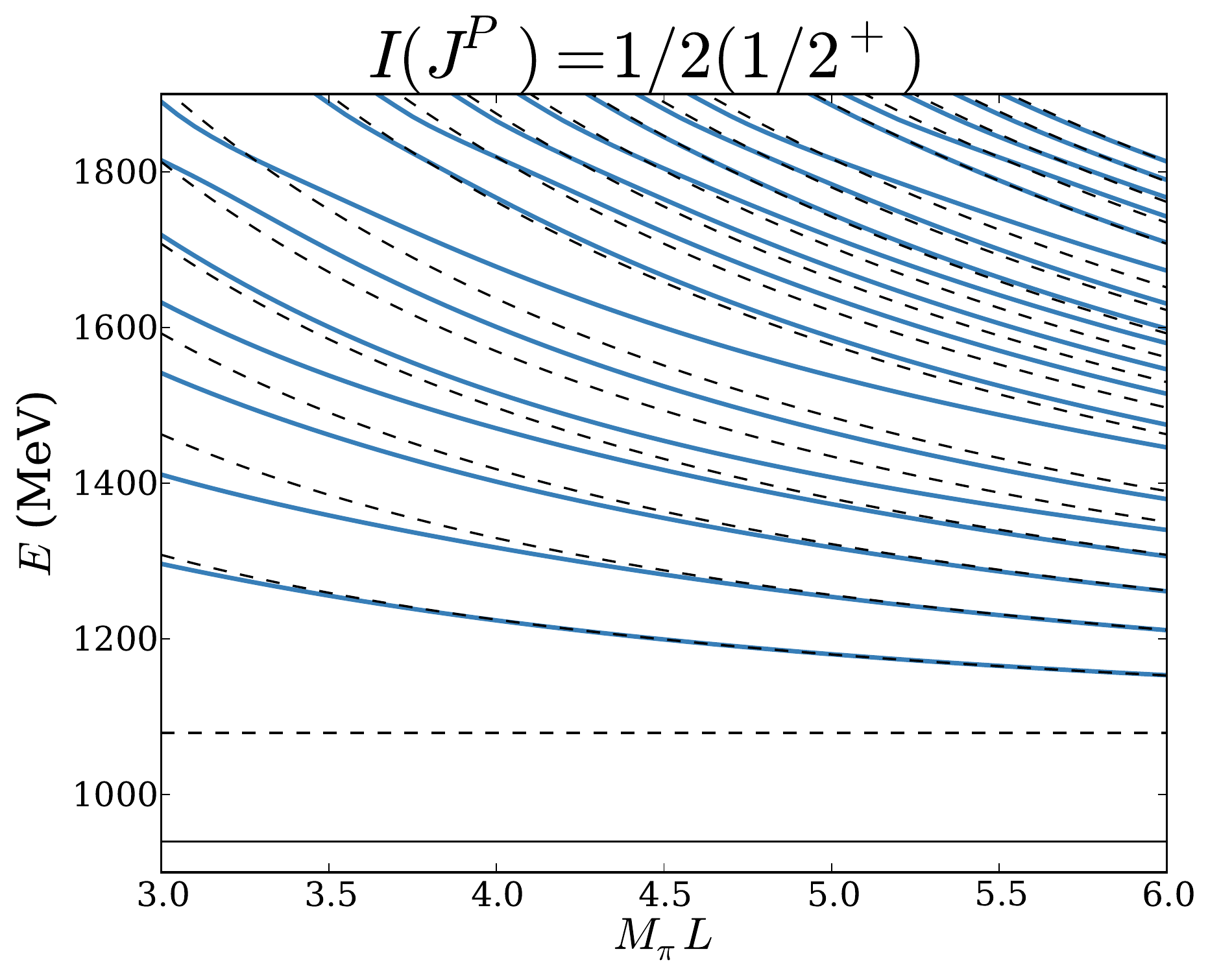}
\caption{Interacting finite-volume $N \pi$ states with $I(J^{P}) = \nicefrac[]{1}{2} (\nicefrac[]{1}{2}^{+})$. The dashed, black curves show the non-interacting energy levels.}
\label{fig:spec}
\end{figure}

Substituting this phase shift into Eq.~(\ref{eq:simplestqc}) we reach the prediction for the two-particle energies, shown in Fig.~\ref{fig:spec}. Note that, {relative to the gap to the single nucleon state}, the shift is relatively small between free- and interacting levels. This means that it makes little difference whether one uses the free or interacting finite-volume spectrum for the values of $\Delta E_n$ that enter $R(T,t)$.\footnote{In this work we do not plot an explicit comparison but, as we comment in Sec.~\ref{sec:contam} below, if one uses LO ChPT for the infinite-volume matrix elements, then the effect of interactions in the energies and Lellouch-L\"uscher factors affects the prediction for $R(T,T/2)$ at the percent level.} Also apparent from Fig.~\ref{fig:spec} is that no avoided level crossing is visible. This is because the Roper resonance is too broad to generate such an effect. It follows that, near the physical point, no direct association between LQCD energies and the resonance can be made and a careful L\"uscher based analysis will be needed to extract resonance properties from LQCD.

\begin{figure}
\includegraphics[scale=0.45]{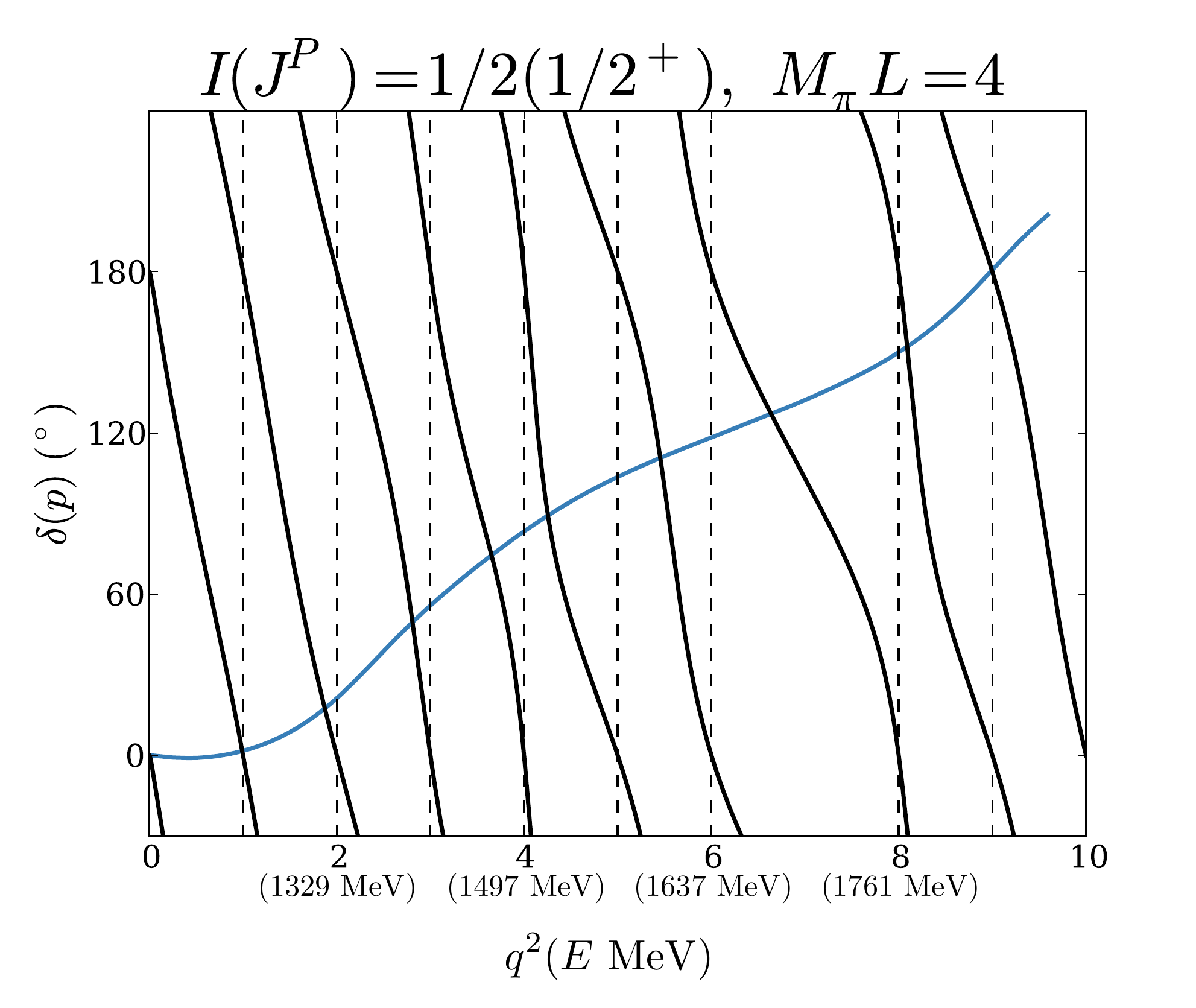}
\vspace{-10pt}
\caption{Experimental scattering phase together with L\"uscher pseudophase curves, $(n \pi - \phi(q))$, for $M_\pi L = 4$. Each intersection gives an interacting level in terms of $q^2$, which can be converted to energy via $E = \sqrt{M_\pi^2 + (2 \pi/L)^2 q^2} + \sqrt{m_N^2 + (2 \pi/L)^2 q^2}$.}
\label{fig:phaseandps}
\end{figure}

To better understand these results consider the form of the pseudophase curves, plotted together with the experimental phase shift for $M_\pi L = 4$ in Fig.~\ref{fig:phaseandps}. The interacting energies, at this $L$ value, are given by the intersections of the curves. This shows that there are universal features for the levels predicted by certain types of phase shifts. In particular, for any phase shift that slowly rises from $0$ to $\pi$, the spectrum is given by a smooth deformation of the free levels. When $\delta(p)$ is near $0$ or $\pi$ the energies coincide with free values. As one follows a given interacting level from high energies to low (by increasing $M_\pi L$) it rises by one free level. This implies that, for any slowly rising phase shift, interacting levels tend to be separated from their neighbors on each side by levels of the free theory. Also, the rise of the phase shift from $0$ to $\pi$ results in exactly one additional energy level relative to the free theory. 

Finally, as we have already stressed above, in this prediction of the energy levels we neglect the effects of crossing three-particle production threshold. Roughly speaking crossing this threshold has two effects. First, three-particle states appear on top of the two-particle states shown in the figure. Second, the positions of all energies are modified relative to those predicted by the two-particle L\"uscher formula. Strictly it does not make sense to distinguish between two- and three-particle states. All finite-volume states will have both two- and three-particle components once the energy exceeds $2 M_\pi + m_N$. However, for sufficiently weak two-to-three couplings, the levels are well described by being two-particle or three-particle like in certain regions, with avoided level crossings occurring whenever a given level changes from one type to the other. The overlap of the interpolator on a given state is also expected to be suppressed when the state has a large three-particle component, possibility with the exception of energies near the Roper. These observations, and the limitation of the formalism available, motivate us to use the effective spectrum plotted in Fig.~\ref{fig:spec} in our study of excited-state contamination.

%%%%%%%%%%%%%%%%%%%%%%%%%%%
%%%%%%%%%%%%%%%%%%%%%%%%%%%
%%%%%%%%%%%%%%%%%%%%%%%%%%%
%%%%%               COEFFICIENTS             %%%%%
%%%%%%%%%%%%%%%%%%%%%%%%%%%
%%%%%%%%%%%%%%%%%%%%%%%%%%%
%%%%%%%%%%%%%%%%%%%%%%%%%%%

\section{Estimating the finite-volume matrix elements}

\label{sec:LL}

In this section we use experimental scattering data, together with LO ChPT and a model to describe $N \pi$ final-state interactions, in order to estimate the finite-volume matrix elements entering $b_n$ and $c_n$. The finite-volume two-particle states, denoted by $\vert n \rangle$, arise from insertions of the identity in Eqs.~(\ref{eq:sd3}) and (\ref{eq:sd2}) and always appear as an outer product of the form $\vert n \rangle \langle n \vert$. This is exactly the structure that is readily accommodated by the generalized Lellouch-L\"uscher formalism, as we now describe.

The original work by Lellouch and L\"uscher gives the relation between a finite-volume matrix element and the $K \to \pi \pi$ decay rate \cite{Lellouch2000}
\begin{equation}
\label{eq:LLresult}
\langle \pi \pi, E_n, L \vert \mathcal H(0) \vert K,L \rangle^2 =   \frac{\vert \mathcal R \vert}{ 2 M_K L^6}     \vert \mathcal H^{\mathrm{out}}  \vert^2 \,,
\end{equation}
where $M_K$ is the kaon mass, $\mathcal H$ is the weak hamiltonian density in position space, $\mathcal H^{\mathrm{out}}  \equiv \langle \pi \pi, \mathrm{out} \vert \mathcal H(0) \vert K \rangle $ is the corresponding infinite-volume matrix element and 
\begin{equation}
\vert \mathcal R \vert =  \frac{p}{16 \pi M_K}    \left [  \frac{\partial}{\partial E} \left( \phi + \delta_{\pi \pi}  \right ) \right ]_{E=M_K}^{-1} \,,
\end{equation}
where $\phi$ is defined in Eq.~(\ref{eq:phidef}) above and $\delta_{\pi \pi}$ is the s-wave phase-shift for elastic pion scattering. The finite-volume states on the right-hand side of Eq.~(\ref{eq:LLresult}) are unit normalized, whereas the infinite-volume states within $\mathcal H^{\mathrm{out}}$ satisfy the standard relativistic normalization. In the derivation of Lellouch and L\"uscher the box size must be tuned so that the two-pion and kaon states are degenerate, $E_n = M_K$.

As has become increasingly clear through various subsequent studies, see for example Refs.~\cite{Lellouch2000, KSS2005, Christ2005, MeyerTimelike2011, HarveyPhotodis, HSmultiLL, BricenoTwoPart2012, BernardUnstable2012, BHWOneToTwo, BHOneToTwoSpin}, the conversion factor, $\vert \mathcal R \vert$, can be understood as a relation between the two-particle states defined in finite and infinite volume. This means that essentially the same relation holds even after relaxing a number of assumptions going into the original derivation. In particular one can define a Lellouch-L\"uscher like relation for operators with generic quantum numbers and with nonzero energy and momentum.%
%%%%%%%%%
% FOOTNOTE
%%%%%%%%%
\footnote{In the context of $K \to \pi \pi$ this relaxes the need to tune the two-pion state to be degenerate. However, if one does not perform this tuning then the resulting infinite-volume matrix element has an energy difference between incoming and outgoing states, and thus looses its straightforward physical interpretation.} %
%%%%%%%%%
% FOOTNOTE
%%%%%%%%%
For products of matrix elements that involve an outerproduct of finite-volume states, the relation can be derived without taking magnitudes of matrix elements \cite{BHWOneToTwo}. Thus, information about the relative sign of transition processes can also be obtained.

In the context of this study, the relevant relation, given in Ref.~\cite{BHOneToTwoSpin}, takes the form
\begin{multline}
\label{eq:genLL}
  \langle 0 \vert \widetilde {\mathcal O}^{}_\beta(0) \vert n \rangle \langle n \vert \widetilde A^{3}_\mu(0) \vert 1 \rangle    = \\      \langle 0 \vert {\mathcal O}_\beta(0) \vert N \pi,  \mathrm{in} \rangle \frac{L^{3/2} \mathcal R(E_n,L)}{\sqrt{2 m_N} } \\ \times \langle N \pi, \mathrm{out} \vert A^{3}_\mu(0) \vert N \rangle   \,,
\end{multline}
where $\mathcal R(E_n,L)$ is a matrix generalization of $\vert \mathcal R \vert $, defined in the following subsection. This is the conversion needed for $b_n$. The analog for $c_{2,n}$ is given by
\begin{multline}
\label{eq:genLL00}
  \langle 0 \vert \widetilde {\mathcal O}^{}_\beta(0) \vert n \rangle \langle n \vert \widetilde {\overline {\mathcal O}}_\alpha(0) \vert 0 \rangle    = \\      \langle 0 \vert {\mathcal O}_\beta(0) \vert N \pi,  \mathrm{in} \rangle L^{3} \mathcal R(E_n,L) \\ \times \langle N \pi, \mathrm{out} \vert \widetilde {\overline {\mathcal O}}_\alpha(0) \vert 0 \rangle   \,.
\end{multline} 

The key limitation of Eqs.~(\ref{eq:genLL}) and (\ref{eq:genLL00}) is that these only hold for $E_n < 2 M_\pi + m_N$. For such two-particle energies the relation is valid up to exponentially suppressed corrections of the form $e^{- M_\pi L}$, but above three-particle threshold a generalized form with three-particle matrix elements is required. As in the previous section, here we again apply the two-particle formalism outside of its region of validity. We expect this to give a qualitative indication of the nature of excited-state contamination, but only by applying a rigorous three-particle formalism can one reach a reliable quantitative prediction, especially in the vicinity of the Roper.

Applying Eqs.~(\ref{eq:genLL}) and (\ref{eq:genLL00}), with $\mathcal R(E_n,L)$ determined using experimental scattering data, it remains only to analyze the matrix elements of the nucleon interpolating operator, $\mathcal O$, and of the axial-vector current, $A_\mu$, in infinite volume. In this way, the details of the finite-volume set-up are factored out. To explain this in detail we find it convenient to assume specific forms for the projectors entering Eqs.~(\ref{eq:C3def}) and (\ref{eq:C2def}). In particular we take $\Gamma = u_+(\textbf 0) \bar u_+(\textbf 0)/m_N$ and $\Gamma'_\mu = \delta_{3\mu} 2 i  \Gamma$, were $u$ and $\overline u$ are standard nucleon spinors, already used in Eq.~(\ref{eq:gAdef}). One then finds
\begin{align}
\label{eq:bnignor}
b_n & = B(E_n) \mathcal C(E_n,L) \mathcal A(E_n) \,, \\
c_{2,n} & =  2   m_N  \omega_N \omega_\pi L^3 B(E_n) \mathcal C(E_n,L) B^\dagger(E_n) \,,
\end{align}
where
\begin{align}
B(E_n) & \equiv  \frac{2i}{   2 \omega_{N} 2 \omega_{\pi} L^3 }  \frac{  \langle 0 \vert \mathcal O_+(0) \vert N \pi, E_n, \mathrm{in} \rangle  e^{- i \delta}   }{  \langle 0 \vert   \mathcal O_+(0)  \vert N  \rangle  }  \,, \\[5pt]
\label{eq:Cdef}
\mathcal C(E_n, L) & \equiv 2 \omega_{\pi} 2 \omega_{N} L^3 e^{ i \delta} \mathcal R(E_n, L) e^{ i \delta} \,, \\[10pt]
\mathcal A(E_n) & \equiv e^{- i \delta} \langle N \pi, E_n, \mathrm{out} \vert A^{a=3}_{\mu=3}(0) \vert N \rangle \,.
\end{align}

Here the first factor, $B(E_n)$, is understood as a row-vector on the $J, \ell, \mu$ index space labeling the two-particle state. It depends on the spin-projected interpolator
\begin{equation}
  {\mathcal O}_+  \equiv \frac{1}{\sqrt{m_N}} \, \bar u_+(\textbf 0) \cdot   {\mathcal O}     \,,
\end{equation}
as well as the kinematic factors  $\omega_\pi$ and $\omega_N$, evaluated at momentum $p$ as defined in Eq.~(\ref{eq:pdef}). The middle factor, $\mathcal C(E_n,L)$, is a matrix on the $J, \ell, \mu$ space. We discuss its definition in detail in the following subsection. Finally $\mathcal A(E_n)$, understood as a column on the same index space, is the infinite-volume axial-vector transition amplitude. 

We comment that all three of the factors entering Eq.~(\ref{eq:bnignor}) are dimensionless, real functions. The latter claim holds due to the diagonal matrices $e^{- i \delta}$ included in the definitions. Here $\delta$ is a diagonal matrix of $N \pi$ scattering phase shifts. For example, the $J=1/2, \ell=1$ entry is plotted in Fig.~\ref{fig:pshifts}. Watson's theorem states that the complex phase of a two-particle matrix element (below three-particle production threshold) is equal to the elastic $N \pi$ scattering phase in the same channel~\cite{Watson}. Thus the phase matrices in the definitions cancel those in the infinite-volume matrix elements. Above three-particle threshold this no longer holds, but in this work we model the matrix elements with a form satisfying this two-particle unitarity constraint. In other words we build in the approximation that Watson's theorem persists above threshold. [See Sec.~\ref{sec:infME} for details.] Similarly the factors of $e^{ i \delta}$ in Eq.~(\ref{eq:Cdef}) cancel the intrinsic phase in $\mathcal R(E_n,L)$ as we show in the next section. This ensures that $\mathcal C(E_n,L)$ is also a real function.

In the following subsection we give the matrix definition of $\mathcal R(E_n, {}L)$ and $\mathcal C(E_n,L)$ and explain that, in the present case, one can truncate these to single entries by applying the same truncation used for the scattering amplitude in the previous section. In Sec.~\ref{sec:predLL} we then use experimental scattering data to calculate the interacting values of $\mathcal C(E_n, L)$. Finally in Sec.~\ref{sec:infME} we use a model, based in LO ChPT supplemented by the experimental scattering data, to estimate both $B(E)$ and $\mathcal A(E)$. We then apply these results in Sec.~\ref{sec:contam}, to give predictions for the excited-state contamination to $g_A$.

\subsection{Reducing the Lellouch-L\"uscher-like relation}

\label{sec:redLL}

We begin this subsection by defining $\mathcal R(E_n, L)$, introduced in Eq.~(\ref{eq:genLL}) above. In this equation the right-hand side should be understood as the product of a column vector $   \langle 0 \vert {\mathcal O}_\beta(0) \vert N \pi,  \mathrm{in} \rangle $, followed by the matrix $\mathcal R(E_n, L)$, followed by a row vector $  \langle N \pi, \mathrm{out} \vert A^{3}_\mu(0) \vert N \rangle$. Each of these quantities is defined on the $J, \ell, \mu$ index space, where the three labels correspond to total angular momentum, orbital angular momentum, and azimuthal total angular momentum respectively. The matrix $\mathcal R(E_n,L)$ is defined by
\begin{equation}
\label{eq:Rintro}
\mathcal R(E_{n}, L) \equiv  \lim_{E \rightarrow  E_{n}} \left[  (E - E_{n}) \frac{1}{F^{-1}(E,  L) + \mathcal M(E)}\right] \,,
\end{equation}
with $\mathcal M$ and $F$ defined in Eqs.~(\ref{eq:MJdef}) and (\ref{eq:FJdef}) respectively. $\mathcal R$ has both on- and off-diagonal elements and, in the context of Eq.~(\ref{eq:genLL}), gives a linear combination of infinite-volume matrix elements that equals a particular finite-volume matrix element. The same matrix structure holds in Eq.~(\ref{eq:bnignor}).

To truncate $\mathcal R$ we first observe that the operator $A^3_3(0)$ acting on the infinite-volume single-nucleon state generates a state which couples to both $J=1/2$ and $J=3/2$. In the corresponding finite-volume matrix element this state couples to two-particle finite-volume states in the $G_1^- = 1/2 \oplus \cdots$ and $H^- = 3/2 \oplus \cdots$ representations. Thus, if we choose the two-particle state to transform in the $G_1^-$ irrep, then the right-hand sides of Eqs.~(\ref{eq:genLL}) and (\ref{eq:bnignor}) will contain one term, depending on the $J=1/2$ two-particle scattering state
\begin{multline}
%\label{eq:genLL}
  \langle 0 \vert \widetilde {\mathcal O}_\beta(0) \vert n, G_1^- \rangle \langle n, G_1^- \vert \widetilde A^3_3(0) \vert 1 \rangle    = \\      \langle 0 \vert {\mathcal O}_\beta(0) \vert N \pi, J=1/2, \mathrm{in} \rangle \frac{L^{3/2} \mathcal R_{J=1/2}(E_n,L)}{\sqrt{2 m_N} } \\ \times  \langle N \pi, J=1/2, \mathrm{out} \vert A^3_3(0) \vert N \rangle   \,.
\end{multline}

Given this truncation we are left only to determine the on-diagonal $J=1/2, \ell=1$ entry of $\mathcal R$. In principle this single entry depends on the full matrix structure of $F^{-1}$ and $\mathcal M$, since they enter via a matrix inverse. However, if we apply the p-wave truncation on $\mathcal M$, as in the previous section, then $\mathcal M$ and $F$ both truncate to single entry matrices. We find \cite{Lellouch2000}
\begin{align}
\mathcal R(E_{n},L ) & =   \left [ \frac{\partial}{\partial E} \left( F^{-1}(E,L) + \mathcal M(E) \right ) \right ]_{E=E_n}^{-1}  \,, \\
& \hspace{-40pt} = -\frac{p}{8 \pi E}   \left [ \sin^2\! \delta \  e^{ 2i \delta }  \frac{\partial}{\partial E} \left( \cot\phi+ \cot\delta \right ) \right ]_{E=E_n}^{-1}    \,, \\
& \hspace{0pt} = \frac{p}{8 \pi E} e^{-2i \delta }  \left [  \frac{\partial}{\partial E} \left( \phi + \delta  \right ) \right ]_{E=E_n}^{-1}    \,,
\label{eq:R1D}
 \end{align}
 where $\delta(p) = \delta_{J=1/2,\ell=1}(p)$, is the $N \pi$ phase shift, shown in Fig.~\ref{fig:pshifts}.

To understand the phase in $\mathcal R$, we recall from Watson's theorem that, at energies where only two-particle elastic scattering can occur, the complex phase of zero-to-two and one-to-two transition amplitudes is given by the two-to-two strong scattering phase~\cite{Watson}. Thus the phase in $\mathcal R$ perfectly cancels the phase in the matrix element
\begin{align}
  e^{-i \delta}   \langle N \pi, \mathrm{out} \vert A^3_3(0) \vert N \rangle \in \mathbb R \,. 
\end{align}

We conclude by discussing the rescaled quantity $\mathcal C(E_n,L)$, defined in Eq.~(\ref{eq:Cdef}). Substituting Eq.~(\ref{eq:R1D}) into the definition and simplifying, we reach
\begin{equation}
\label{eq:Cres}
\mathcal C(E,L)  =  4 \pi^2 q^3   \left( q  \frac{\partial \phi}{\partial q} + p  \frac{\partial \delta}{\partial p}  \right )^{-1} \,.
\end{equation}
In next section we will also be interested in the non-interacting limit, and thus define
\begin{equation}
\label{eq:Cres2}
\mathcal C^{\mathrm{NI}}(q^2)  \equiv  4 \pi^2 q^2  \left(   \frac{\partial \phi}{\partial q}   \right )^{-1} \,,
\end{equation}
where $q \equiv p L/(2 \pi)$ was already introduced above. Note that in Eqs.~(\ref{eq:Cres}) and (\ref{eq:Cres2}) we have implicitly extended the definition of $\mathcal C(E,L)$ to all energies. As is clear from Eqs.~(\ref{eq:genLL}) and (\ref{eq:bnignor}), the quantity only has physical \mbox{meaning} when evaluated at the energies of the finite-volume spectrum. However, understanding the continuous form of the function is useful for predicting how $\mathcal C(E_n,L)$ will vary with the strength of the particle interaction. 

\begin{figure}
\begin{center}
%75
\vspace{0pt}
\includegraphics[scale=0.45]{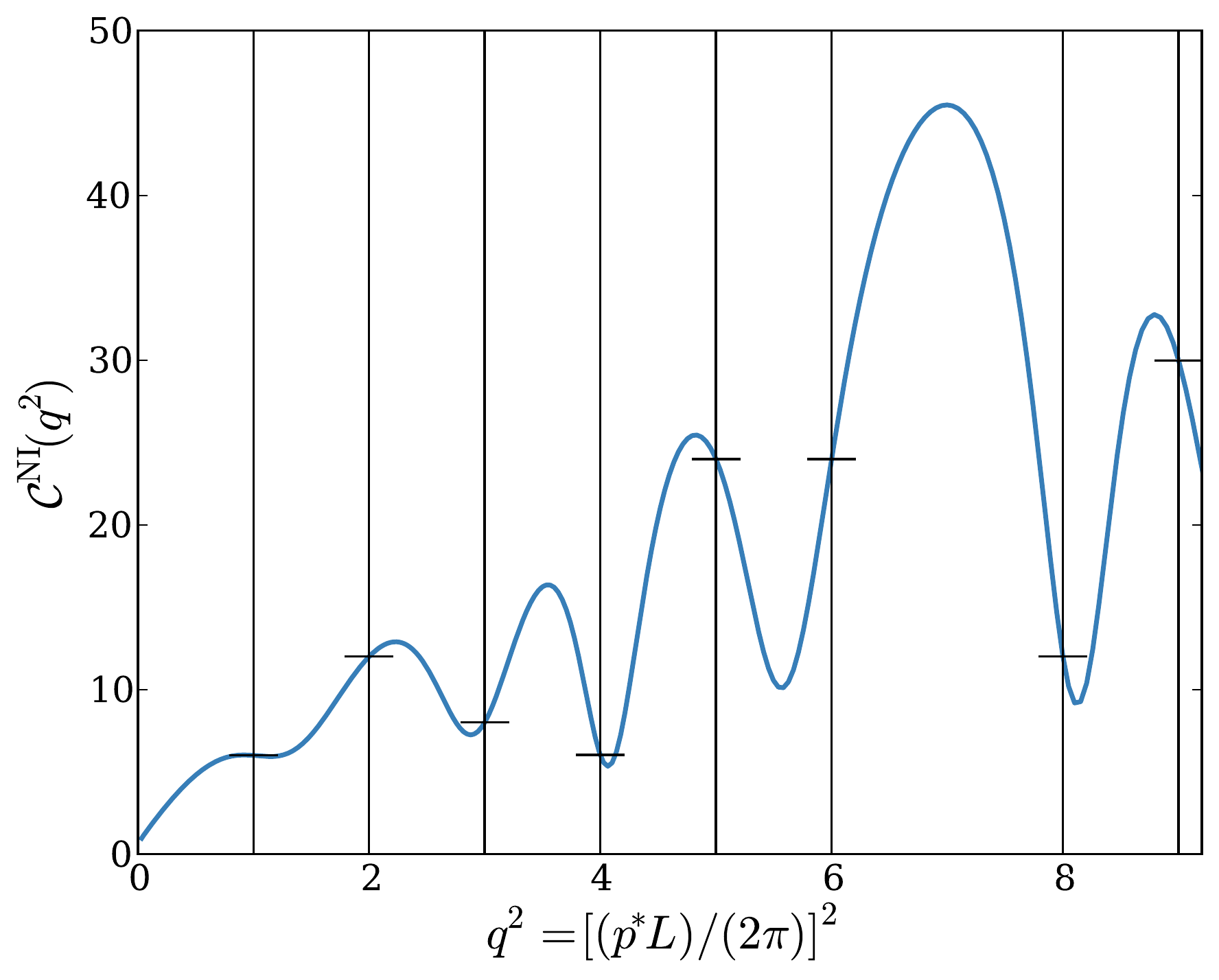}
\end{center}
\caption{Non-interacting Lellouch-L\"uscher curve. This curve, defined in Eq.~(\ref{eq:Cres2}), only has a clear physical meaning at the non-interacting finite-volume energies, indicated by vertical lines. Here it coincides with the degeneracy of the finite-volume state, $\nu_n$. Considering the form of the curve everywhere is useful for understanding the effect of interactions, as shown in Fig.~\ref{fig:intLL}.}
\label{fig:nonintLL}
\end{figure}

\subsection{\label{sec:predLL}Predicting the Lellouch-L\"uscher factors}

In this section we give numerical predictions for the values of $\mathcal C(E_n,L)$ and in doing so also build some intuition about the meaning of this quantity. 

We begin with the non-interacting version, $\mathcal C^{\mathrm{NI}}$. This is plotted in Fig.~\ref{fig:nonintLL} as a function of the dimensionless squared momentum, $q^2$. The energies for which this curve has physical meaning correspond to $q^2 = \textbf n^2$ with $\textbf n \in \mathbb Z^3$. At these values our rescaled Lellouch-L\"uscher factor takes on particularly simple values
\begin{equation}
\label{eq:Cfree}
\mathcal C^{\mathrm{NI}}(n) = \nu_n \,, 
\end{equation}
where $\nu_n$ is the degeneracy of the $n$th state, equivalently the number of integer vectors that satisfy $\textbf n^2 = n$. The first few values of $\nu_n$ are given in Table~\ref{tab:deg}. These degeneracies are also indicated by the horizontal tick marks crossing each vertical line in Fig.~\ref{fig:nonintLL}. 

\begin{table}
\begin{tabular}{c|c|c|c|c|c|c|c|c|c|c|c|c|c|c|c|c}
\hline \\[-9pt]
\hline
$n$         & 0 & 1 & 2 & 3 & 4 & 5 & 6 & 7 & 8 & 9 & 10 & 11 & 12 & 13 & 14 & 15 \\
$\nu_n$ & 1& 6& 12& 8& 6& 24& 24& 0& 12& 30& 24& 24& 8& 24& 48& 0 \\
\hline 
 \end{tabular}
\caption{Degeneracies of states with $q^2 = n$.}
\label{tab:deg}
\end{table}

\begin{figure}
\begin{center}
%75
\vspace{-25pt}
\includegraphics[scale=0.45]{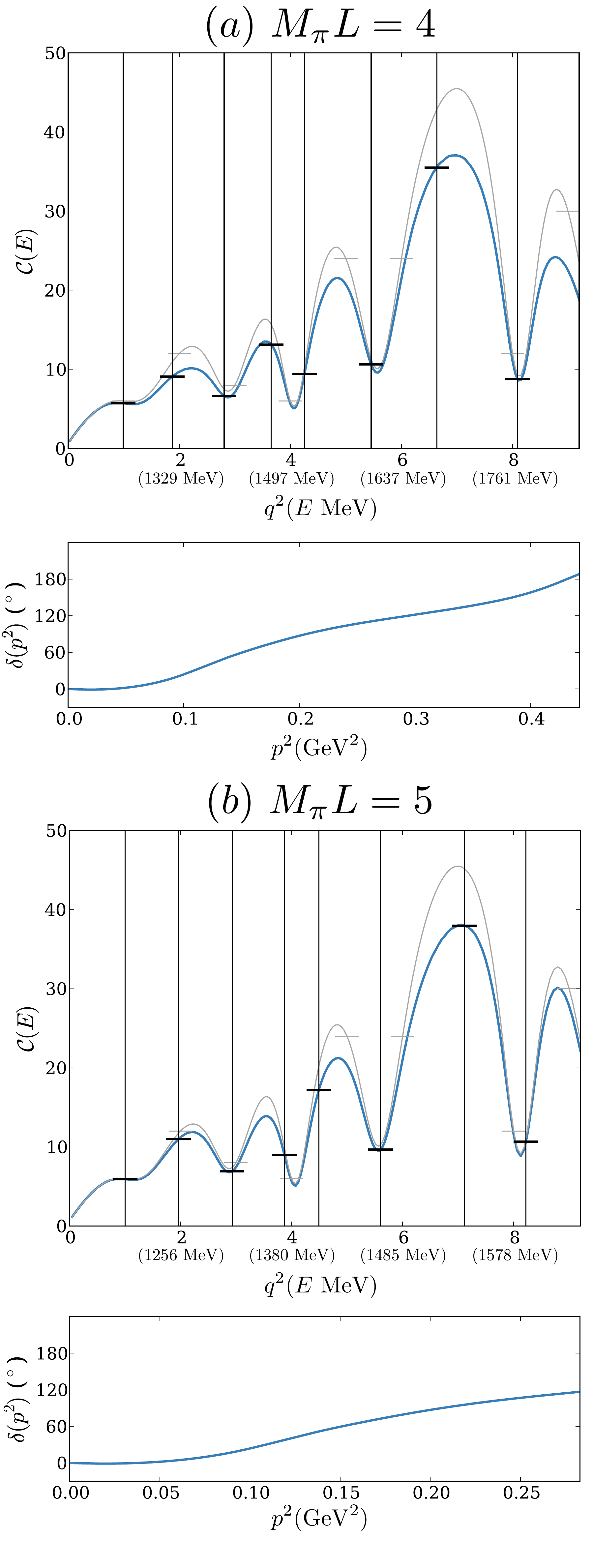}
\vspace{-25pt}
\end{center}
\caption{These plots summarize the effect of interactions on $\mathcal C$ for (a) $M_\pi L=4$ and (b) $M_\pi L=5$ as a function of $q^2 = (p L)^2/(2 \pi)^2$. We also indicate the two-particle energies corresponding with certain $q^2$ values. The interacting curve (blue) is lower than the non-interacting (gray) due to a shift proportional to $d \delta(p)/dp$ in the inverse. For ease of comparison we have plotted the phase shift over the same range for each $M_\pi L$ value. However, the most important effect for the interacting value of $\mathcal C$ is the shift in energies. Since the curve is rapidly oscillating, these shifts result in large discrepancies between the interacting and non-interacting $\mathcal C$ values. }
\label{fig:intLL}
\end{figure}

We next turn to the interacting values of $\mathcal C(E_n,L)$, plotted in Fig.~\ref{fig:intLL}. In contrast to $\mathcal C^{\mathrm{NI}}$, the factor in the interacting case depends independently on $E$ and $L$. Here we plot $\mathcal C(E,L)$ as a function of $q^2$ at fixed $M_\pi L = 4$ [Fig.~\ref{fig:intLL}(a)] and $M_\pi L = 5$ [Fig.~\ref{fig:intLL}(b)]. Note that two effects are important in the shift from non-interacting to interacting systems. First, the curve characterizing the Lellouch-L\"uscher factors is reduced, due to the addition of a term proportional to $d \delta/dp$ in the inverse. Comparing the light gray and blue curves, we see that this has a relatively small numerical effect. Second, the physically relevant values on the curve (the finite-volume energies) are shifted to new locations. Interestingly, this second shift has a large effect for both $M_\pi L=4$ and $M_\pi L=5$ (assuming physical pion masses). In particular, the contribution to $R(T,t)$ from the seventh excited state is significantly enhanced due to the large value of $\mathcal C(E_n,L)$. 

\subsection{\label{sec:infME}Modeling the matrix elements}

We now turn to the remaining building blocks entering the definitions of $b_n$ and $c_n$: the overlap factor $B(E_n)$ and the axial-vector transition amplitude $\mathcal A(E_n)$. 

As already mentioned above, the matrix elements within $B(E)$ depend on the details of the interpolator, $\mathcal O_+$, and cannot be constrained using experimental data. The precise definition of $\mathcal O_+$ depends on the set-up of a particular lattice calculation, and the only clear defining property is that it has the quantum numbers to annihilate a single nucleon. In a variational approach, this operator is designed to reduce the size of the two-particle matrix elements shown in Eq.~(\ref{eq:genLL}). Thus by choosing a sufficiently large basis one can make this contribution arbitrarily small.%
%%%%%%%%%
% FOOTNOTE
%%%%%%%%%
\footnote{Given that $\mathcal O$ can be optimized to minimize excited-state contamination, the universal result found by Refs.~\cite{BrianGA,BarTwoPoint,BarGA} may seem surprising. We stress, however, that the ChPT predictions only hold for local and smeared operators. The results thus suggest that $N \pi$ interpolating operators are probably required to systematically reduce the overlap to the low-lying excited states.} 
%%%%%%%%%
% FOOTNOTE
%%%%%%%%%

By contrast, $\mathcal A(E)$ is a physical matrix element that can in principle be accessed in a scattering experiment. In fact, since we are focusing on the case where both the nucleon and the $N \pi$ state are at rest, parity guarantees that the corresponding matrix element with a vector current must vanish. Thus we can equally well think of $\mathcal A(E)$ as the matrix element of the left-handed (vector-minus-axial) current. The kinematics of $\mathcal A(E)$ correspond to a process such as $p \, \pi^- \to p \, W^- \to p \, e^- \, \bar \nu^e$, in which the current is evaluated at time-like four-momentum. Such a transition is difficult to extract in experiment and the present data is insufficient to constrain the amplitude. We also note that the value of this matrix element at space-like momenta is of great experimental relevance for determining QCD effects in neutrino-nucleon scattering, $\nu_\ell p \to p \pi^+ \ell^-$.%
%%%%%%%%%
% FOOTNOTE
%%%%%%%%%
\footnote{See for example Ref.~\cite{AalvarezRuso2014}.}
%%%%%%%%%
% FOOTNOTE
%%%%%%%%%

In this work we rely on LO ChPT, together with a model that incorporates the experimental $N \pi$ scattering data, in order to gain insight into the behavior of both the interpolator overlap $B(E)$ and the axial-vector transition amplitude $\mathcal A(E)$. Beginning with $B(E)$, we first give the expression predicted by LO covariant baryon ChPT, derived in the appendix
\begin{equation}
\label{eq:BnChPT}
  B_{\mathrm{ChPT}}(E) = \frac{\sqrt{3}}{4 \sqrt{2} f_\pi   \omega_{\pi }   \omega_{N} L^3  }  \left    (\frac{\omega_N}{m_N} - 1 \right)^{1/2}  (1 - \bar g_A) \,,
  \end{equation}
where
\begin{equation}
\label{eq:gAbardef}
 \bar g_A \equiv g_A \frac{E_n+m_N}{E_n-m_N} = g_A \frac{\omega_N + \omega_\pi +m_N}{\omega_N + \omega_\pi -m_N}   \,,
\end{equation}
is a convenient shorthand introduced in Ref.~\cite{BarTwoPoint}, and $f_\pi=93{\,\mathrm{MeV}}$ is the pion decay constant. This result predicts the part of $b_n$ that depends on $\mathcal O$ to be negative and have magnitude $\sim10^{-3}$. In addition, as was already pointed out in Refs.~\cite{BrianGA,BarTwoPoint,BarGA}, the leading-order prediction is independent of the details of the interpolator used. 
Again, we stress that this only holds for a three-quark
interpolating field and, in particular, does not apply to any
interpolator built from multi-particle operators, for example $N \pi$- or
$N \pi \pi$-like operators.

Eq.~(\ref{eq:BnChPT}) is expected to break down at higher energies, when
next-to-leading-order ChPT corrections become important. For instance,
if $\mathcal O$ is optimized to couple to a single-nucleon state then,
depending on the nature of the Roper, the overlap of the operator with
two-particle states may also be enhanced in the vicinity of the
resonance. This enhancement is not visible in LO ChPT.

To model the effect of the Roper we first consider the Bethe-Salpeter equation for the two-particle matrix element within $B(E)$
\begin{multline}
  \langle 0 \vert \mathcal O_+(0) \vert N \pi, \mathrm{in} \rangle =   \langle 0 \vert \mathcal O_+(0) \vert N \pi, \mathrm{in} \rangle_{2\mathrm{PI}} \ + \\ \int \frac{d^4 k}{(2 \pi)^4} \langle 0 \vert \mathcal O_+(0) \vert N \pi, \mathrm{in} \rangle_{2\mathrm{PI}} \, \Delta(k) S(P-k) \, i \mathcal M(E,k) \,,
   \label{eq:BetheSalpteterBn}
\end{multline}
where the subscript $2\mathrm{PI}$ refers to the sum of all diagrams that are two-particle irreducible in the s-channel. To reach the full matrix element, the 2PI quantity must be attached, via a two-particle loop, to the two-to-two scattering amplitude, $\mathcal M$. In the loop both the pion [$\Delta (k)$] and nucleon [$S(P-k)$] propagators should be fully dressed and evaluated at the off-shell momenta sampled by the integral. The scattering amplitude is also sampled at off-shell momenta.

We cannot evaluate this quantity without making a number of approximations. First we evaluate the $k^0$ integral, approximating the matrix element and $\mathcal M$ to have no important analytic structure, so that we need only encircle the poles in the two-particle loop. This gives
\begin{multline}
  \langle 0 \vert \mathcal O_+(0) \vert N \pi, \mathrm{in} \rangle =   \langle 0 \vert \mathcal O_+(0) \vert N \pi, \mathrm{in} \rangle_{2\mathrm{PI}} \ + \\ \int \frac{d^3 \textbf k}{(2 \pi)^3} \langle 0 \vert \mathcal O_+(0) \vert N \pi, \mathrm{in} \rangle_{2\mathrm{PI}} \, \\ \times \left[ -\frac{u(\textbf k) \overline u(\textbf k)}{2 \omega_N 2 \omega_\pi  (   E - \omega_N - \omega_{\pi} + i \epsilon)}  + \mathcal S(\vec k) \right ]  \mathcal M(E,k) \,,
\end{multline}
where $\mathcal S$ is a smooth function below three-particle production threshold. If we assume the dominant contribution comes form the first term, and drop the smooth part then we are left with a contribution in which both the matrix element and the scattering amplitude are projected on shell. We find
\begin{multline}
  \langle 0 \vert \mathcal O_+(0) \vert N \pi, \mathrm{in} \rangle =   \langle 0 \vert \mathcal O_+(0) \vert N \pi, \mathrm{in} \rangle_{2\mathrm{PI}}  \\ \times [1 +  \mathcal I_R(E)    \mathcal M(E) ] \,,
\end{multline}
where
\begin{equation}
\mathcal I_R(E)  =   \frac{i p}{8 \pi E} - \mathrm{PV} \! \int_{R} \! \! \frac{d^3 \textbf k}{(2 \pi)^3}   \frac{1}{2 \omega_N 2 \omega_\pi  (   E - \omega_N - \omega_{\pi}  )}  \,,
\end{equation}
and where PV indicates a principal-value pole prescription. The subscript $R$ indicates that this loop integral requires a regulator, an artifact that has been introduced by our approximations. In this work we choose to regulate by subtracting the integrand evaluated at threshold
\begin{multline}
\mathcal I(E)  \equiv   \frac{i p}{8 \pi E}  - \mathrm{PV} \! \int  \! \! \frac{d^3 \textbf k}{(2 \pi)^3}   \frac{1}{2 \omega_N 2 \omega_\pi   }  \\ \times \left[\frac{1}{  E - \omega_N - \omega_{\pi}}  - \frac{1}{  m_N + m_\pi - \omega_N - \omega_{\pi}}  \right ] \,.
\end{multline}
This subtraction is motivated by the observation that the second term in Eq.~(\ref{eq:BetheSalpteterBn}) should not play a role at low energies. Note also that the on-shell restriction projects the scattering amplitude down to its $I(J^{P}) = \nicefrac[]{1}{2} (\nicefrac[]{1}{2}^{+})$ component.

To complete the construction of our model we use the fact that the diagrams in the LO ChPT calculation of $B(E)$ are two-particle irreducible, and thus also give the leading-order contribution to the 2PI restriction of this quantity. This leads us to define
\begin{equation}
B(E,\gamma) \equiv   \mathrm{Re} \left [ e^{- i \delta} B_{  \mathrm{ChPT}}(E) \left[ 1 + \gamma \mathcal I(E) \mathcal M_{}(E)  \right ]  \right ] \,,
\end{equation}
where we have introduced the free parameter $\gamma$ to partially compensate the {\em ad hoc} procedure that lead us to this expression. Here we have also included the phase factor $e^{- i \delta}$ that is needed to cancel the phase that appears in the two-particle matrix element.%
%%%%%%%%%
% FOOTNOTE
%%%%%%%%%
\footnote{This simple phase structure is only strictly valid below three-particle production threshold. In the present model we are using the elastic form for the two-particle scattering amplitude and this has the consequence that the phase is preserved also above production threshold. This is consistent with the neglect of three-particle states in the L\"uscher quantization condition.} %
%%%%%%%%%
% FOOTNOTE
%%%%%%%%%
To evaluate both the $e^{- i \delta}$ factor and the scattering amplitude $\mathcal M$ we use the experimentally determined $N \pi$ scattering phase, plotted in Fig.~\ref{fig:pshifts}. Note also that we must discard a small imaginary part that arises in this model.

\begin{figure}
\begin{center}
%75
\vspace{20pt}
\hspace{-15pt}\includegraphics[scale=0.45]{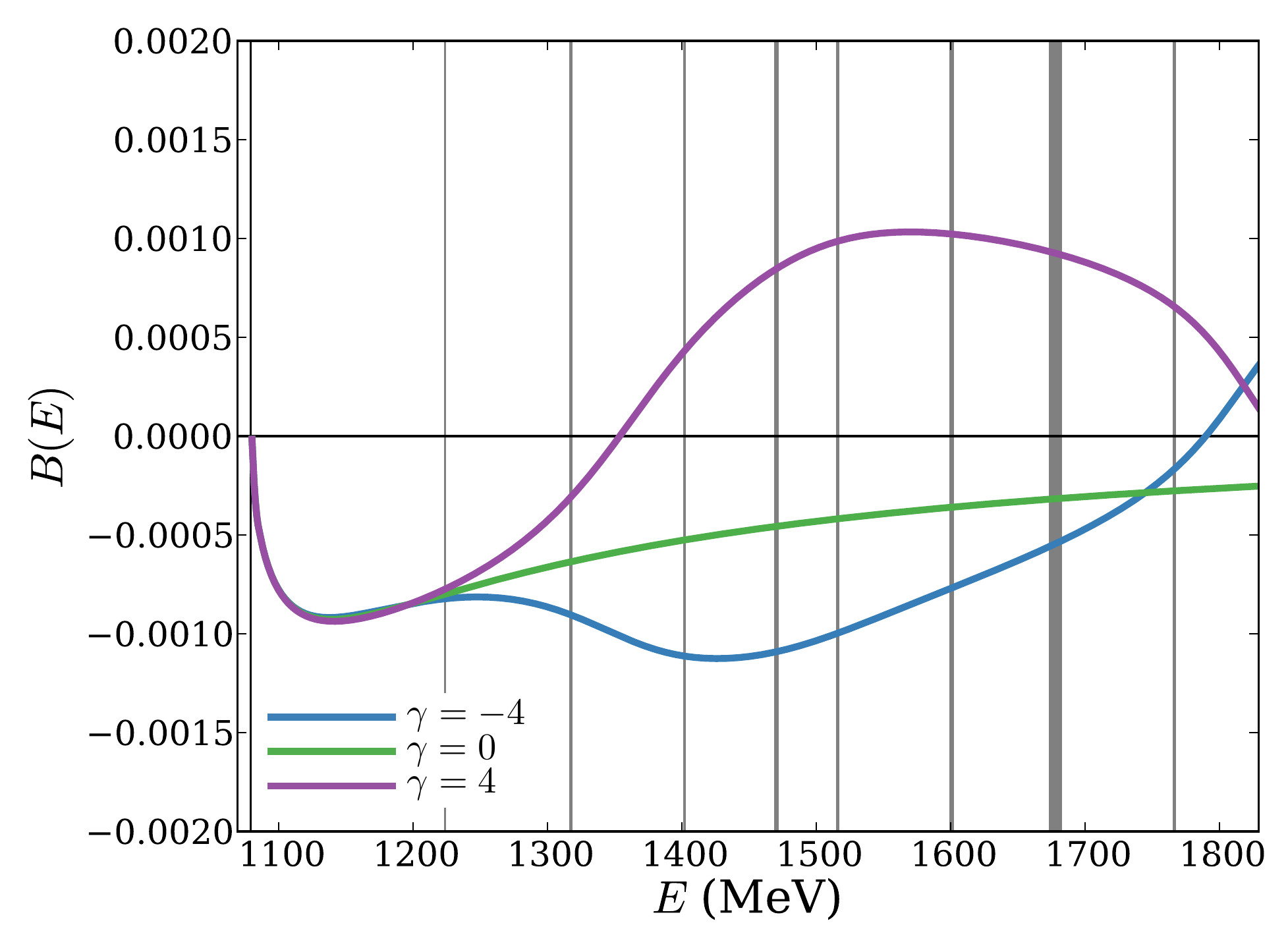}
\vspace{-15pt}
\end{center}
\caption{Three different scenarios for the overlap factor $B(E)$ with $\gamma$ values of $-4$ (lowest), $0$ (middle) and $4$ (highest). The vertical lines indicate the finite-volume energies in a box of size $M_\pi L=4$. The thickness of these lines is proportional to the value of $\mathcal C(E_n,L)$ and thus indicates how the state is weighted in the excited-state contamination.}
\label{fig:Bnmodels}
\end{figure}

In the case of $\gamma=0$ we omit the $e^{- i \delta}$ phase factor and thus recover the leading order ChPT result. For $\gamma>0$ the rising phase shift causes the matrix element to flip sign roughly in the region of the Roper, with the energy value at the node dependent on the specific value chosen for $\gamma$. For $\gamma<0$ the sign of the ChPT prediction is preserved and a peak is observed in the vicinity of the Roper. Past this energy range the matrix element can flip sign, but for $\gamma<-3$ the crossing is well outside the relevant energy range. In Fig.~\ref{fig:Bnmodels} we plot the energy dependence predicted by various values of $\gamma$.

We now turn to the matrix element of the axial-vector current, $\mathcal A(E)$. As we derive in the appendix, the LO ChPT prediction for this quantity is
\begin{multline}
 \mathcal A(E)_{\mathrm{ChPT}} =   \frac{ \sqrt{m_N}(\omega_{N} - m_N)^{1/2} }{ 2 \sqrt{6} f_\pi  }     \\
  \times  \left[4-\frac{8}{3} g_A \left(\overline g_A-\frac{g_A M_\pi^2}{  4 \omega_{\pi} m_N-2 M_\pi^2 }\right)\right] \,.
\label{eq:AxialChPT}
\end{multline}
To estimate this quantity beyond ChPT we apply the same model used for the overlap factor
\begin{equation}
 \mathcal A(E ,\alpha) = \mathrm{Re} \left [ e^{- i \delta }  \mathcal A(E)_{\mathrm{ChPT}}  [1 + \alpha \mathcal I(E) \mathcal M(E)] \right ]\,.
\end{equation}
As with $B(E,\gamma)$, $\alpha = 0$ gives the LO ChPT prediction, $\alpha<0$ preserves the sign of the matrix element and enhances the magnitude near the resonance, and $\alpha>0$ gives a zero-crossing roughly in the range of the resonance. Also as with $B(E,\gamma)$, we include the phase factor needed to cancel the phase in the matrix element, and discard a small imaginary contribution that arises as an artifact of our model. For $\alpha=0$ the phase factor is not included.  

\begin{figure}
\begin{center}
%75
\vspace{19pt}
\hspace{-10pt} \includegraphics[scale=0.45]{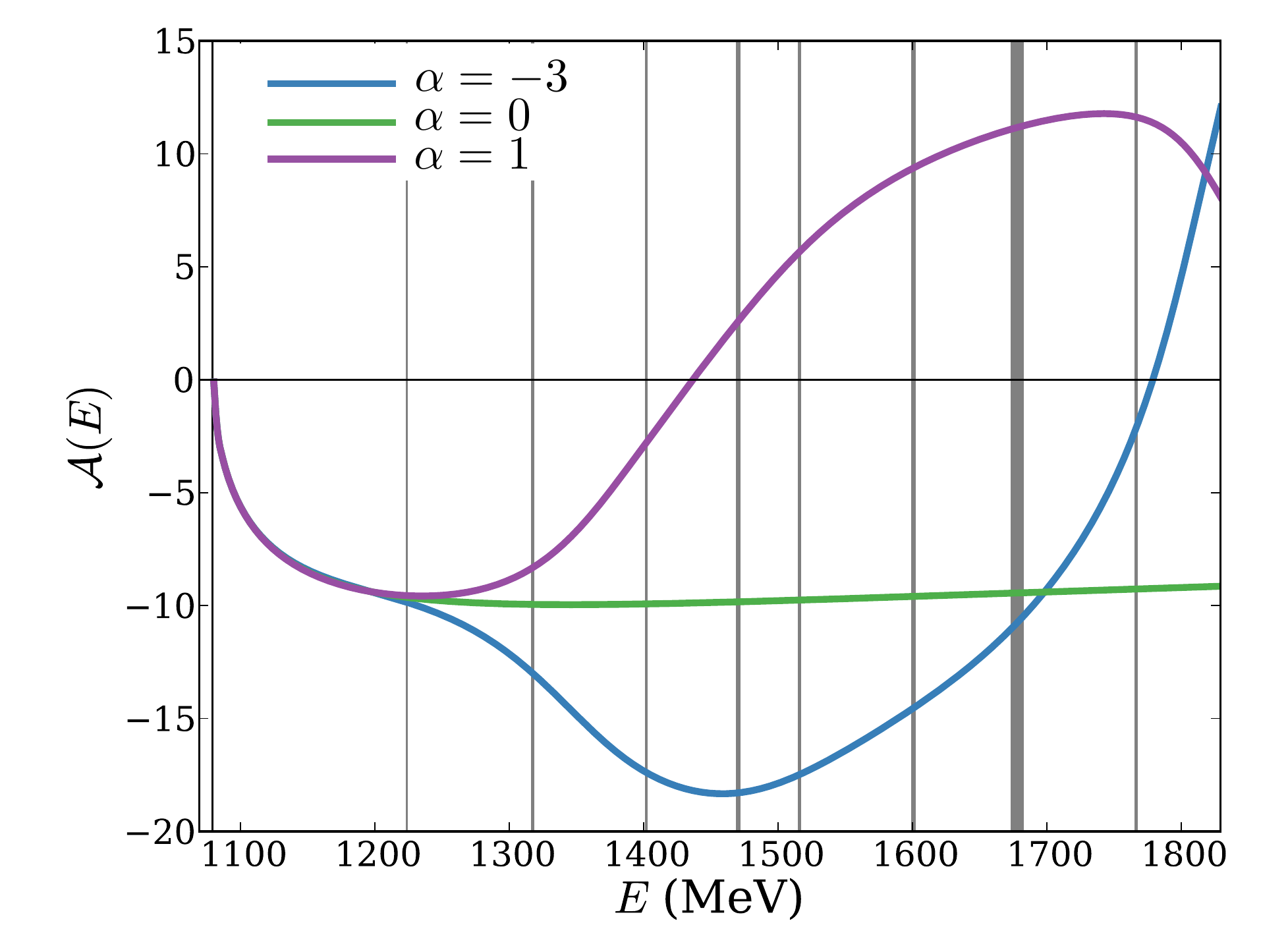} \hspace{10pt}
\vspace{-25pt}
\end{center}
\caption{Three different scenarios for the axial matrix element $\mathcal A(E)$ with $\alpha$ values of $-3$ (lowest), $0$ (middle) and $1$ (highest). As in Fig.~\ref{fig:Bnmodels}, the vertical lines indicate finite-volume energies for $M_\pi L=4$ with line thickness proportional to $\mathcal C(E_n,L)$.}
\label{fig:Amodels}
\end{figure}

In Fig.~\ref{fig:Amodels} we plot the energy dependence of the \mbox{axial} transition amplitude for various choices of $\alpha$. Note that we restrict attention to a range of $\alpha$ that is smaller than that considered for $\gamma$. The LO ChPT prediction for $B(E)$ has a magnitude that decreases with energy whereas $\mathcal A(E)$ is nearly constant at higher energies. This has the consequence that varying $\alpha$ over a given range has a larger effect than varying $\gamma$. We choose the parameters such that the models have a maximum magnitude roughly within a factor of two of the maximum predicted by LO ChPT.

We stress again that, unlike with the overlap factor, the functional form of $\mathcal A(E)$ is independent of the lattice set-up and is of direct experimental relevance. In this study we are particularly interested in whether $\mathcal A(E)$ has a node (zero crossing) at some energy, as predicted by the $\alpha>0$ models. Interestingly, such a node is observed in the CLAS scattering data for $e  p \to e' \pi^+ \pi^- p'$ in their analysis of the electromagnetic transition amplitude as a function of photon virtuality, $Q^2$ \cite{RoperCLAS}. This node is not directly relevant for $\mathcal A(E)$ because (1) it concerns the electromagnetic transition and (2) it is for space-like momenta. It is nonetheless interesting to note that such crossings are observed.

Given that LO ChPT predicts the same sign for $B(E)$ and $\mathcal A(E)$, and thus a positive value for $b_n = B(E_n) \mathcal C(E_n,L) \mathcal A(E_n)$, and given also that the curvature in LQCD correlator data indicates important contributions from states with $b_n<0$, we postulate that a node in $\mathcal A(E)$ might provide a reasonable explanation for the apparent discrepancy. More generally one can identify four basic scenarios: (i) neither $B(E)$ nor $\mathcal A(E)$ cross zero in the relevant energy range, (ii) both cross zero, (iii) only the overlap factor $B(E)$ has a node, or (iv) only the transition amplitude $\mathcal A(E)$ has a node. The first two cases lead to positive excited-state contamination and thus fail to describe present day numerical LQCD correlator data. The third and fourth scenarios can both explain the empirically observed excited state-contamination, as we explain in the next section.

\section{\label{sec:contam}Estimating the excited-state contamination}

We are now ready to combine the results of the previous section to estimate the ratio $R(T,t)$ and the excited-state contamination to $g_A$.

\begin{figure}
%\begin{center}
%75
\vspace{20pt}
\hspace{-25pt} \includegraphics[scale=0.48]{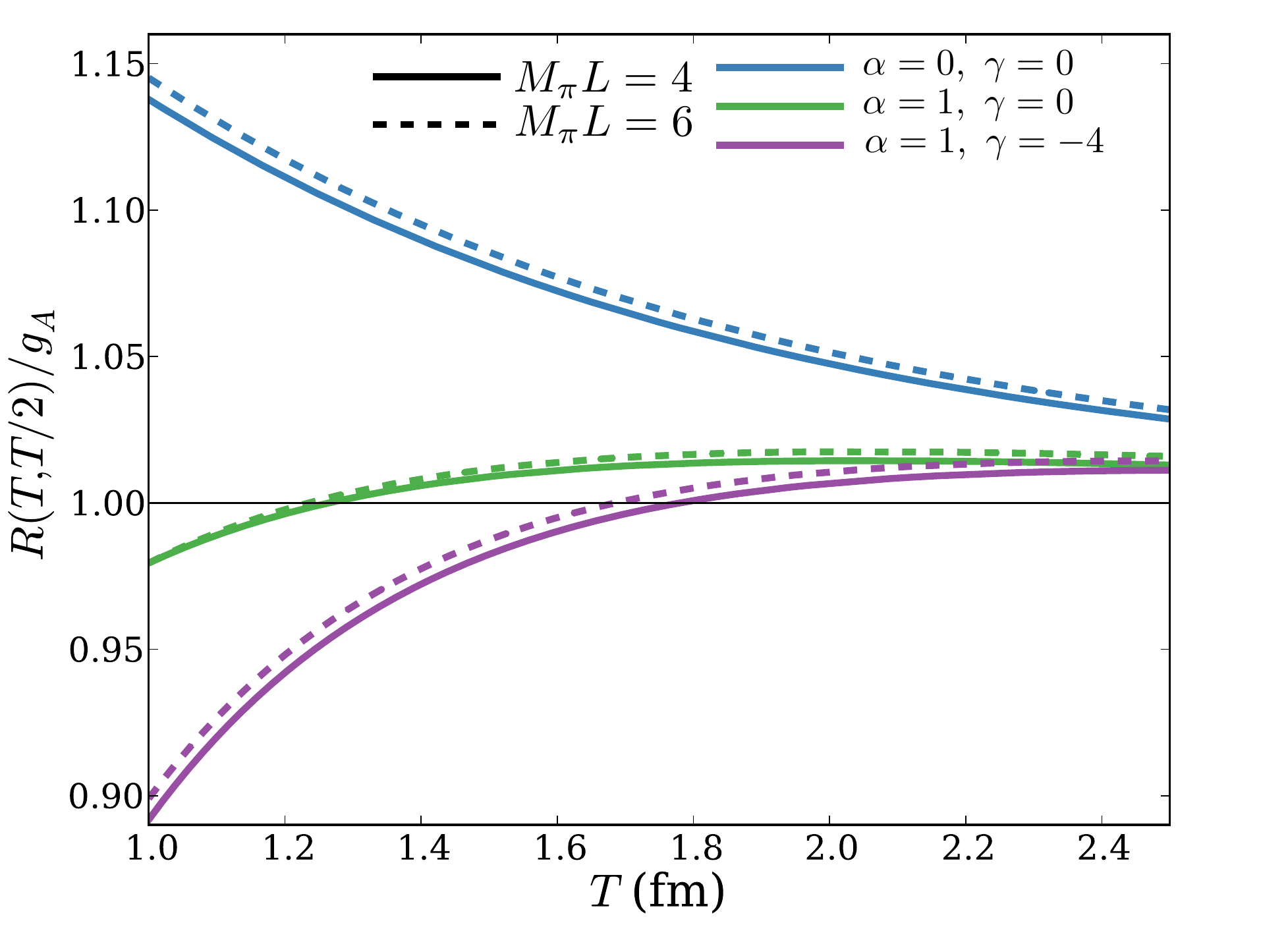} \hspace{145pt}

\vspace{-20pt}

%\end{center}
\caption{Excited-state contamination for various values of $\alpha$ [parametrizing $\mathcal A(E)$] and $\gamma$ [parametrizing $B(E)$] for $M_\pi L = 4$ (solid) and $M_\pi L = 6$ (dashed). The top pair of curves shows the leading order ChPT prediction, but with the interacting values for $\mathcal C(E_n,L)$, the middle pair shows the leading order ChPT value for $B(E)$ together with the $\alpha=1$ (zero-crossing) scenario for $\mathcal A(E)$. Finally, the bottom pair shows the result of setting $\alpha=1$ together with $\gamma=-4$. Of the parameter sets considered here this choice most closely reproduces the observed LQCD correlator data. This lowest curve compares favorably, for example, with the $M_\pi = 190 \mathrm{\,MeV}$ and $M_\pi L = 3.9$ ensemble in Ref.~\cite{EigoAMA2016}.}
\label{fig:esc1}
\end{figure}

In Fig.~\ref{fig:esc1} we show values for the ratio $R(T,T/2)/g_A$, given three different scenarios for the matrix elements. In each case we show the results for both $M_\pi L = 4$ (solid lines) and $M_\pi L=6$ (dashed lines). To provide comparable predictions, in both cases we sum all excited states up to an energy of 1800 MeV. In the case of $M_\pi L = 4$ this corresponds to the first 8 excited states and for $M_\pi L = 6$ to the first 18. In both cases we find that this number of states is both necessary and sufficient to estimate the saturated value of $R(T, T/2)/g_A$ within a few percent, for the models considered. We also see that the excited state contaminations from the two different volumes are in very good agreement.

The highest pair of curves in Fig.~\ref{fig:esc1} shows the prediction from LO ChPT, but with the interacting values of the Lellouch-L\"uscher factors. The excited-state contamination here is comparable to that given in Fig.~5 of Ref.~\cite{BarGA}, in particular the result of that reference that includes ten excited states. Thus we find that, if one uses LO ChPT for the overlap and the axial-vector matrix element, then the effect of interactions on the energies and Lellouch-L\"uscher factors leads to a small (percent level) correction to the predicted value of $R(T,T/2)$.

As is also stressed in Ref.~\cite{BarGA}, including excited states beyond the first few requires sampling the LO ChPT predictions outside their expected region of validity. For example, for $M_\pi L=4$ it is well motivated to trust the first two excited states. These have a gap of $\sim 100 \, \mathrm{MeV}$ to the Roper and, as the latter is not included in the ChPT prediction, it is reasonable to require a separation from this state. 
We thus infer that one can only predict $R(T,T/2)$ using LO ChPT for source-sink separations large enough that the first two states dominate. For physical pion masses this may mean separations of $T > 2 \mathrm{\,fm}$.

The middle pair of curves in Fig.~\ref{fig:esc1} is the prediction from combining the LO ChPT prediction for the overlap, $B(E,\gamma=0)$, with the zero-crossing model for the axial-current transition amplitude, $\mathcal A(E,\alpha=1)$. The curve is very flat because there are large cancellations between the positive contributions from lower states and the negative contributions from higher-energy states. 

The lowest pair of curves in Fig.~\ref{fig:esc1} gives the scenario most consistent with observed LQCD correlators. This follows from combining $\mathcal A(E,\alpha=1)$ and $B(E,\gamma=-4)$ [see again Figs.~\ref{fig:Bnmodels} and \ref{fig:Amodels}]. The negative contribution from the higher excited states overpowers the positive contribution from the first few.

\begin{figure}
\begin{center}
%75
\vspace{20pt}
\includegraphics[scale=0.45]{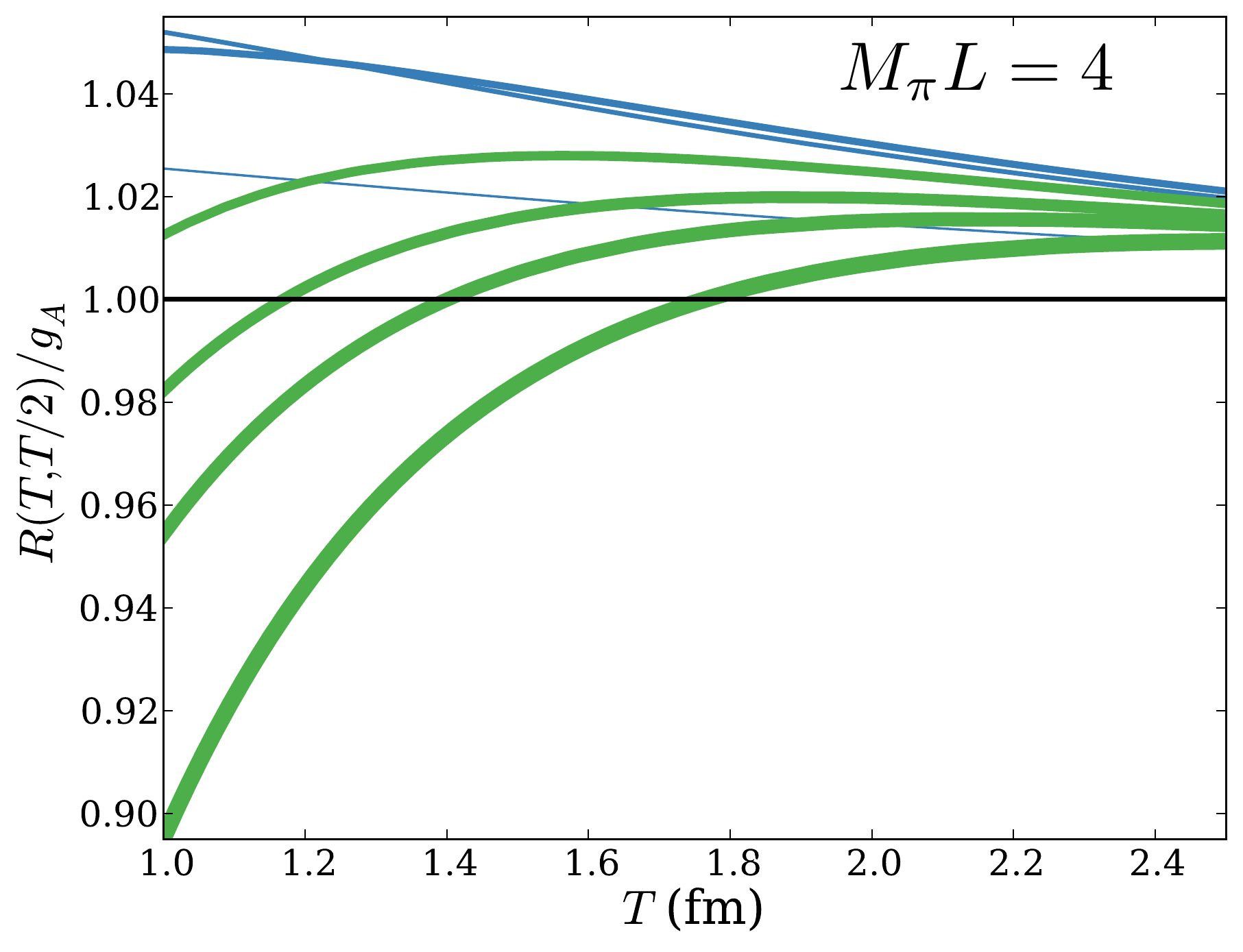}
\end{center}
\caption{Contribution of individual excited states in the $\alpha=1$, $\gamma=-4$ scenario, for $M_\pi L=4$. The seven curves show the excited-state contamination predicted by summing the contributions from the first state through the $n$th state. The thickness of a given curve is proportional to the number of states included in the sum. The three blue curves show the result from including the first (lowest blue), first two (middle blue) and first three (highest blue) excited states. The fourth excited state is the first with $b_n<0$ so that the predicted contamination falls once this state is included (highest green curve). The set of green curves then indicates the sum up to the fourth state (highest green) to the sum up to the seventh state (lowest green). The effects of including additional states beyond the seventh are negligible so that the lowest green curve gives a good indication of the full excited state contamination predicted by this model. }
\label{fig:esc2}
\end{figure}

In Fig.~\ref{fig:esc2} we show the importance of higher excited states in the $\mathcal A(E,\alpha=1)$ and $B(E,\gamma=-4)$ scenario. In particular, for $M_\pi L=4$, we find that the contamination predicted by summing fewer than seven states differs significantly from the saturated curve. In particular the seventh excited state has a significant contribution due to the large value of $\mathcal C(E_n,L)$.

%%%%%%%%%%%%%%%%%%%%%%%%%%%
%%%%%%%%%%%%%%%%%%%%%%%%%%%
%%%%%%%%%%%%%%%%%%%%%%%%%%%
%%%%%                CONCLUSION              %%%%%
%%%%%%%%%%%%%%%%%%%%%%%%%%%
%%%%%%%%%%%%%%%%%%%%%%%%%%%
%%%%%%%%%%%%%%%%%%%%%%%%%%%

\section{Conclusion}

In this work we have studied the excited-state contamination in LQCD correlators used to extract the nucleon axial charge, $g_A$. Combining various finite-volume formalisms with experimental scattering data, LO ChPT and a model for the infinite-volume matrix elements, we find that the excited-state behavior empirically observed in lattice correlators can be reproduced by postulating a sign change in the infinite-volume axial-vector transition amplitude, $\langle N \pi, \mathrm{out} \vert A \vert N \rangle$. Such nodes are observed experimentally in other transition amplitudes, but the data is insufficient to make a definitive statement about the quantity at hand.

Our findings additionally indicate that a large number of finite-volume excited states, including those at energies around the Roper resonance, give important contributions in the ratios used to access $g_A$. This is based on mild assumptions about how the nucleon interpolators couple to states near the Roper and on the observation that the Lellouch-L\"uscher factors, governing the relation between finite- and infinite-volume states, can be significantly enhanced. 

The results presented here serve to further emphasize the great importance of using optimal interpolators to minimize the coupling to excited states. Based on numerical LQCD calculations in the meson sector, the most promising approach seems to be the variational method, in which a large basis of operators is used to disentangle the excited states. The situation will also be improved by further advances in improving the signal-to-noise ratio in nucleon correlators. 

Finally we emphasize that the nature of excited-state contamination depends heavily on the quantity under consideration. Indeed for many quantities, for example average $x$, the LQCD data indicates positive excited-state contamination \cite{ETMAvx2016}. For this observable the same overlap factor $B(E)$ appears, but a different transition-amplitude arises due to the differing current insertion. The observed, positive excited-state contamination can be accommodated if we suppose the matrix element has the same sign as the axial vector at low-energies and does not cross zero in the relevant energy window.  

Another interesting example is the iso-singlet octet axial-vector. ChPT predictions indicate that matrix elements of this current should be highly suppressed relative to those of the iso-triplet studied here. The iso-singlet suffers from other sources of systematic uncertainty, in particular quark-disconnected diagrams. Given the potential severity of excited-state contaminations, it will be interesting to compare the systematic error budgets for these quantities as methods on both sides improve.

\acknowledgements{We thank J. Green, T. Harris, G. von Hippel, P. Junnarkar, D. Mohler, D. Robaina, H. Wittig and all our colleagues in the Mainz lattice
group for helpful discussions, encouragement and support. We thank Oliver B\"ar for very helpful comments on the first version of this manuscript.}

\appendix

\section{ChPT}

In this appendix we summarize the various ChPT results used in the main text. We begin with the calculation of the two-point correlator, defined via
\begin{equation}
C_2(T)  \equiv  L^{-3} \sum_{n} \   \langle 0 \vert \widetilde  {\mathcal O}_+ \vert n \rangle  \langle n \vert   \widetilde {\overline {\mathcal O}}_+ \vert 0 \rangle e^{- E_n T} \,,
\end{equation}
where 
\begin{align}
\widetilde  {\mathcal O}_+ & = \frac{1}{\sqrt{m_N}} \, \bar u_+(\textbf 0) \cdot \widetilde  {\mathcal O} \,, \\
\widetilde  {\overline {\mathcal O}}_+ & = \frac{1}{\sqrt{m_N}} \,  \widetilde {\overline {\mathcal O}}   \cdot u_+(\textbf 0)  \,.
\end{align}

Working in time-momentum perturbation theory
Ref.~\cite{BarTwoPoint} found
\begin{equation}
\label{eq:C2Tresult}
C_2(T) = 2 \vert \tilde \alpha \vert^2  e^{- m_N T} \bigg [ 1 +  \sum_{n=1}^\infty  c_{2,n} e^{- \Delta E_n T}  \bigg ] \,,
\end{equation}
where
\begin{align}
\label{eq:c2result}
c_{2,n} &  \equiv   \frac{3 \nu_n }{16 \omega_\pi f_\pi^2 L^3  }  \left ( 1 - \frac{m_N}{\omega_N}  \right )  (1 - \overline g_A)^2  \,, \\
 \bar g_A &  \equiv g_A \frac{\omega_N + \omega_\pi + m_N}{\omega_N + \omega_\pi - m_N} \,.
\end{align}
Here we have repeated the definition of $\bar g_A$, given in Eq.~(\ref{eq:gAbardef}) above, but with $\omega_N + \omega_\pi$ in place of $E_n$. This is equivalent given the definition of the individual energies
\begin{equation}
\omega_N \equiv \sqrt{m_N^2 + p^2} \,, \ \ \ \ \omega_\pi \equiv \sqrt{m_\pi^2 + p^2} \,,
\end{equation}
where $p$ is defined via
\begin{equation}
E_n \equiv \sqrt{m_N^2 + p^2} +  \sqrt{m_\pi^2 + p^2} \,.
\end{equation}
These definitions are needed to extend the results away from the non-interacting finite-volume energy levels. Eq.~(\ref{eq:C2Tresult}) also depends on $\tilde \alpha$, a low-energy coefficient that was also introduced in Ref.~\cite{BarTwoPoint}. This coefficient enters the relation between the lattice interpolator and the ChPT fields
\begin{equation}
\mathcal O  = \tilde \alpha \left [\begin{pmatrix} p \\ n   \end{pmatrix} +  \frac{i}{2f_\pi} \gamma_{5}   \begin{pmatrix} \pi^0 & \sqrt{2} \pi^+ \\ \sqrt{2} \pi^-  & - \pi^0 \end{pmatrix} \begin{pmatrix} p \\ n   \end{pmatrix}  \right ]\,.
\end{equation}

To cross check the result of Ref.~\cite{BarTwoPoint} we apply the extension of the Lellouch-L\"uscher approach to the relevant finite-volume matrix elements. From the discussion in the main text follows
\begin{align}
c_{2,n} & = \frac{1}{2 \vert \tilde \alpha \vert^2 L^3}   \   \langle 0 \vert \widetilde {\mathcal O}_+ \vert n  \rangle  \langle n  \vert   \widetilde{\overline {\mathcal O}}_+ \vert 0 \rangle \,, \\
& \hspace{-15pt} =  \frac{\nu_n}{8 \vert \tilde \alpha \vert^2   \omega_{\pi }   \omega_{N} L^3 }      \langle 0 \vert {\mathcal O}_+(0) \vert N \pi, \mathrm{in} \rangle   \langle N \pi, \mathrm{out} \vert \overline {\mathcal O}_+(0) \vert 0 \rangle \,.
\label{eq:c2LL}
\end{align}
Here we have used the non-interacting value for the Lellouch-L\"uscher factor as is appropriate for a leading-order calculation. The two-particle asymptotic states on the second line are projected to definite isospin $I=1/2, m_I = 1/2$ definite total angular momentum $J=1/2, \mu=1/2$ and definite parity $P=+$. This implies that orbital angular momentum is restricted to the p-wave. 

\begin{figure}
\begin{center}
%75
\vspace{0pt}
\includegraphics[scale=0.25]{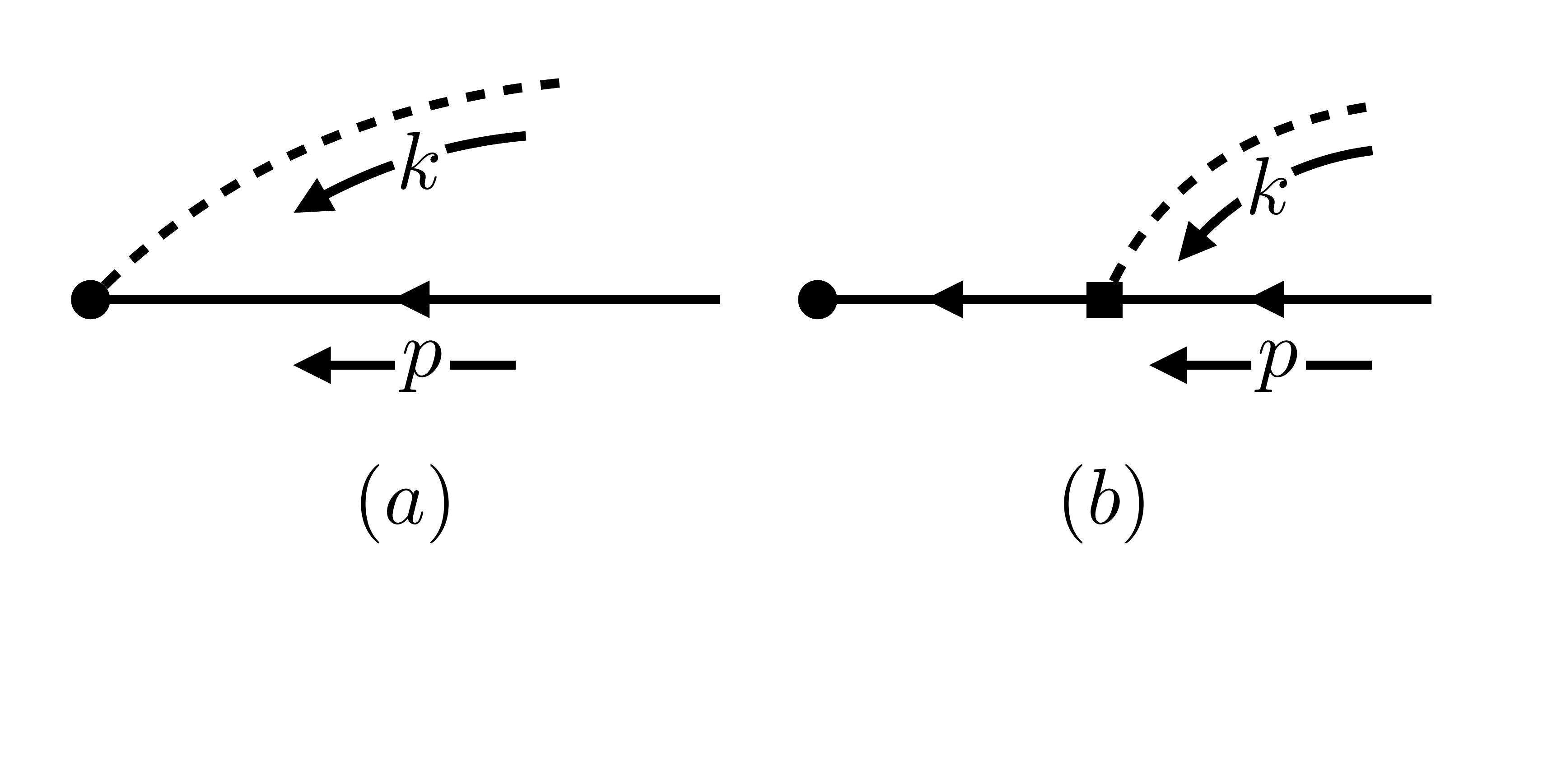}
\end{center}
\vspace{-50pt}
\caption{Feynman diagrams contributing to $\langle 0  \vert \mathcal O_+ \vert N \pi \rangle$ at leading order in ChPT. The filled circle indicates the interpolating field and the square the interaction vertex. The solid and dashed lines represent pions and nucleons, respectively. Here we think of time flowing from right to left so that the diagrams directly correspond with the mathematical expressions.}
\label{fig:interpdiags}
\end{figure}

To reach Eq.~(\ref{eq:c2result}), one must calculate the infinite-volume matrix elements in Eq.~(\ref{eq:c2LL}) at tree level. The relevant two diagrams are shown in Fig.~\ref{fig:interpdiags}. The first of these diagrams, Fig.~\ref{fig:interpdiags}(a), does not include any interactions from the Lagrangian but instead arises from the $N \pi$ contribution in the ChPT expression for the lattice interpolator \cite{BarTwoPoint}
\begin{align}
\mathcal O & \supset \tilde \alpha \frac{i}{2f_\pi} \gamma_{5} \begin{pmatrix} 1 & 0 \end{pmatrix} \begin{pmatrix} \pi^0 & \sqrt{2} \pi^+ \\ \sqrt{2} \pi^-  & - \pi^0 \end{pmatrix} \begin{pmatrix} p \\ n   \end{pmatrix} \,, \\
& =  \tilde \alpha \frac{i}{2f_\pi} \gamma_5 \big  [ \pi^0 p + \sqrt{2} \pi^+ n \big ] \,.
\end{align}
This is then combined with the definite isospin state%
%%%%%%%%%
% FOOTNOTE
%%%%%%%%%
\footnote{Note here that both terms have the same sign, whereas the standard Clebsch-Gordan coefficients have opposite signs. This is because the $\overline p$ and $\overline n$ fields transform in the conjugate $SU(2)$ and a relative minus enters in the similarity rotation to regular $SU(2)$.} %
%%%%%%%%%
% FOOTNOTE
%%%%%%%%%
\begin{equation}
\vert N \pi, \mathrm{in} \rangle =   \bigg [\frac{1}{\sqrt{3}} \pi^0 \overline p    + \sqrt{\frac23}   \pi^- \overline     n  \,    \bigg ] \vert 0 \rangle \,,
\end{equation}
to reach
\begin{multline}
  \langle 0 \vert {\mathcal O}_+(0) \vert N \pi, +,  \mathrm{in} \rangle   \supset  \\  \sqrt{3}  \tilde \alpha   \frac{1}{\sqrt{m_N}} \overline u_+(\textbf 0)    \frac{i}{2f_\pi} \gamma_5  e^{ i \eta \hat p \cdot \vec K }   u_+(\textbf 0) \,.
\end{multline}
Here we have written the right-most spinor, evaluated at the nucleon momentum $p$, as a boosted zero-momentum spinor. We have introduced $\eta = \sinh^{-1}(p/m_N)$ as the rapidity of the nucleon, and $K^i =  -i \gamma^0 \gamma^i/2$ as the generator of the boost. In this appendix we use the Minkowski metric (including Minkowksi gamma matrices), following the conventions of Ref.~\cite{srednicki}. As indicated by the $+$ label on the $N\pi$ state (and on the right-most spinor), this is the result in the case that the incoming nucleon is spin up. 

To simplify we substitute 
\begin{equation}
\label{eq:boostmatrix}
e^{ i \eta \hat p \cdot \vec K } = 
  \cosh (\eta/2)  + \sinh (\eta/2)   \hat p_i \gamma^0 \gamma^i  \,.
\end{equation}
Since $ \overline u_+(\textbf 0) \gamma_5 u_+(\textbf 0) = 0$ only the $\sinh$ term contributes, giving
\begin{multline}
  \langle 0 \vert {\mathcal O}_+(0) \vert N \pi, +,  \mathrm{in} \rangle   \supset    i \tilde \alpha \frac{\sqrt{3 m_N}  }{f_\pi} \sinh (\eta/2)  \hat p_z  \,, \\
=    \frac{1}{\sqrt{3}} i \tilde \alpha  \frac{ \sqrt{3} }{\sqrt{2} f_\pi }  (\omega_N - m_N)^{1/2}\sqrt{4 \pi} Y^*_{10}(\hat p) \,.
\end{multline}
In the second step we used 
\begin{equation}
\sinh \left[ \frac{1}{2} \sinh^{-1} \frac{p}{m_N} \right ]    =   \left [ \frac{ \sqrt{m_N^2+p^2 } - m_N }{2 m_N} \right ]^{1/2} \,,
\end{equation}
and reexpressed the $p_z$ dependence in terms of the p-wave spherical harmonic.

When the incoming nucleon is spin down the spinors pick off a different combination of gamma matrices leading to a different combination of the components of $\hat p$. The result is
\begin{multline}
 \langle 0 \vert {\mathcal O}_+(0) \vert N \pi, -,  \mathrm{in} \rangle   \supset  \\  - \sqrt{\frac23} i \tilde \alpha
\frac{ \sqrt{3} }{\sqrt{2}f_\pi }  (\omega_N - m_N)^{1/2}\sqrt{4 \pi} Y^*_{11}(\hat p) \,.
\end{multline}

Next performing the orbital angular-momentum projection (simply by dropping the $\sqrt{4 \pi}Y^*_{1m}$ factor) and forming the definite $J$ combination
\begin{equation}
|  1/2, 1/2 \rangle = -\frac{1}{\sqrt{3}}   \, \vert + \rangle  \vert 1,0 \rangle +\sqrt{\frac23} \, \vert - \rangle  \vert 1,1 \rangle \,,
\end{equation}
we conclude the final contribution from Fig.~\ref{fig:interpdiags}(a)
\begin{equation}
  \langle 0 \vert {\mathcal O}_+(0) \vert N \pi, \mathrm{in} \rangle  \supset  - i \tilde \alpha \frac{\sqrt{3}}{\sqrt{2} f_\pi} (\omega_N - m_N)^{1/2}   \,.
 \end{equation}
 
We now turn to Fig.~\ref{fig:interpdiags}(b). This diagram arises from the single nucleon contribution to the interpolator together with the $\overline N N \pi$ coupling in the Lagrangian
\begin{equation}
\mathcal L \supset  \frac{g_A}{2 f_\pi} \overline N \gamma^\mu \gamma_5 \sigma^a             N       \partial_\mu \pi^a   \,.
\end{equation}
Inferring the momentum-space Feynman rules and calculating the spin-up matrix element, we find
\begin{multline}
  \langle 0 \vert {\mathcal O}_+(0) \vert N \pi, +,  \mathrm{in} \rangle   \supset    \sqrt{3}  \tilde \alpha   \frac{1}{\sqrt{m_N}} \overline u_+(\textbf 0)     \\ \times  i \frac{(\omega_N + \omega_\pi) \gamma^0 + m_N}{(\omega_N + \omega_\pi)^2 - m_N^2 }   \left( i \frac{g_A}{2f_\pi} (-i \slashed k) \gamma_5 \right ) \\   e^{ i \eta \hat p \cdot \vec K }   u_+(\textbf 0) \,.
\end{multline}
We again substitute Eq.~(\ref{eq:boostmatrix}), but in this case both the cosh and sinh terms contribute giving
\begin{multline}
  \langle 0 \vert {\mathcal O}_+(0) \vert N \pi, +,  \mathrm{in} \rangle   \supset  -   i \tilde \alpha \frac{\sqrt{3 m_N} g_A  }{f_\pi } \frac{1 }{ \omega_N + \omega_\pi - m_N} \\ \times \bigg [  \cosh(\eta/2)  p + \sinh(\eta/2) \omega_\pi     \bigg ]  \hat p_3 \,.
\end{multline}
Applying standard identities for the trigonometric functions one finds
\begin{multline}
  \langle 0 \vert {\mathcal O}_+(0) \vert N \pi, +,  \mathrm{in} \rangle   \supset \\  -   i   \frac{\tilde \alpha}{\sqrt{2} f_\pi }   g_A   (\omega_N - m_N)^{1/2}  \frac{\omega_N + \omega_\pi + m_N}{\omega_N + \omega_\pi - m_N}  \sqrt{4 \pi} Y^*_{10}(\hat p) \,, \\
=   \frac{1}{\sqrt{3}} i \tilde \alpha  \frac{ \sqrt{3} }{\sqrt{2} f_\pi } (- \bar g_A)  (\omega_N - m_N)^{1/2}   \sqrt{4 \pi} Y^*_{10}(\hat p) \,.
 \end{multline}
 
 In words, the contribution to the spin-up state from Fig.~\ref{fig:interpdiags}(b) is just given by the result for \ref{fig:interpdiags}(a) multiplied by $(- \bar g_A)$. The same relation holds for the spin-down state so that the full result for the $J=1/2$ matrix element is
\begin{align}
\label{eq:c2instateME}
  \langle 0 \vert {\mathcal O}_+(0) \vert N \pi, \mathrm{in} \rangle  & =  - i \tilde \alpha \frac{\sqrt{3}}{\sqrt{2} f_\pi} (\omega_N - m_N)^{1/2}  (1 - \bar g_A) \,, \\
    \langle N \pi, \mathrm{out} \vert \overline {\mathcal O}_+(0) \vert 0 \rangle  & =  i \tilde \alpha^* \frac{\sqrt{3}}{\sqrt{2} f_\pi} (\omega_N - m_N)^{1/2}  (1 - \bar g_A) \,.
 \end{align}
Substituting these results into Eq.~(\ref{eq:c2LL}) we recover Eq.~(\ref{eq:c2result}).

\bigskip

We now turn to the three-point function and in particular the ratio used to extract $g_A$. Applying our specific choices of $\Gamma$ and $\Gamma'$ to the definition of $b_n$ we deduce
\begin{align}
b_n  & =   2 i \frac{    \langle 0 \vert \widetilde {\mathcal O}_+ \vert n   \rangle \langle n \vert \widetilde A^3_3 \vert N \rangle  }{    \langle 0 \vert \widetilde {\mathcal O}_+ \vert N \rangle   }    \,, \\
 & =  2 i  \frac{\nu_n}{4     \omega_{\pi }   \omega_{N} L^3  }    \frac{    \langle 0 \vert {\mathcal O}_+(0) \vert N \pi, \mathrm{in}   \rangle \langle N \pi, \mathrm{out} \vert A^3_3(0) \vert N \rangle  }{    \langle 0 \vert {\mathcal O}_+(0) \vert N \rangle   }  \,, \\
 \begin{split}
  & = \frac{\nu_n}{4  \sqrt{m_N}    \omega_{\pi }   \omega_{N} L^3  }    \frac{\sqrt{3}}{\sqrt{2} f_\pi} (\omega_N - m_N)^{1/2}  (1 - \bar g_A) \\ 
  & \hspace{100pt} \times    \langle N \pi, \mathrm{out} \vert A^3_3(0) \vert N \rangle 
  \end{split}\,.
  \label{eq:bLL}
\end{align}
Here we have again applied the Lellouch-L\"uscher formalism and have also substituted Eq.~(\ref{eq:c2instateME}).

\begin{figure}
\begin{center}
%75
\vspace{0pt}
\includegraphics[scale=0.25]{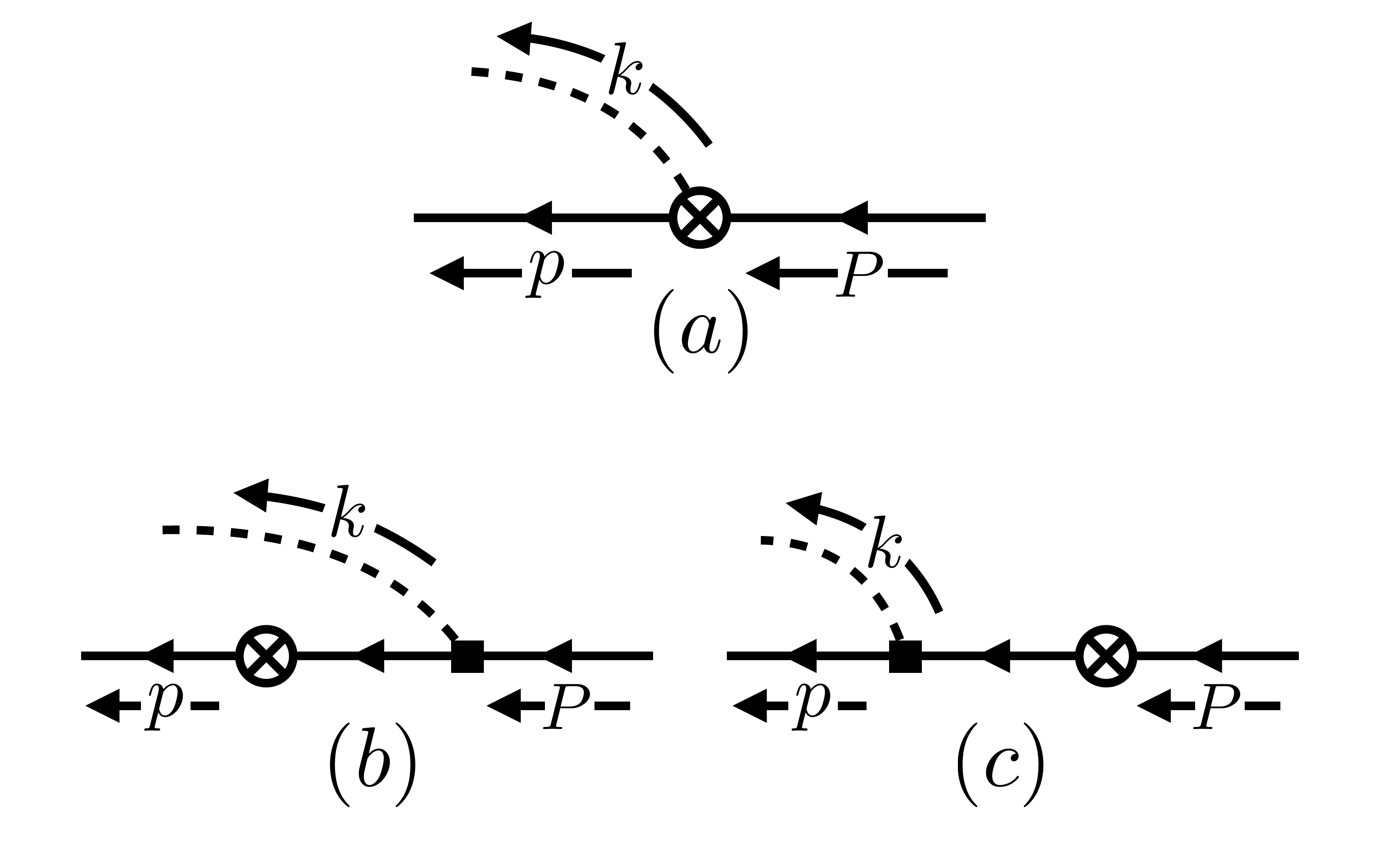}
\end{center}
\vspace{-10pt}
\caption{Feynman diagrams contributing to $\langle N \pi  \vert A_3^3 (0) \vert N \rangle$ at leading order in ChPT. The crossed circle indicates the current insertion and the solid and dashed lines represent nucleons and pions, respectively.}
\label{fig:diags2}
\end{figure}

To complete the calculation it remains only to substitute the result for the infinite-volume matrix element $ \langle N \pi, \mathrm{out} \vert A^3_3(0) \vert N \rangle $. Again this must be evaluated at tree level by calculating the three diagrams shown in Fig.~\ref{fig:diags2}. These diagrams include various contributions from the ChPT expression for the axial-vector current
\begin{equation}
 A^{a}_3 = - i g_A \overline N \gamma^3 \gamma_5 T^a N + i \frac{1}{f_\pi} \epsilon^{abc} \pi^b \overline N \gamma^3 T^c N - i   f_\pi \partial^3 \pi^a \,.
\end{equation}
Here we are using the Euclidean convention for the axial-vector current as is used throughout the main text. However, we are expressing it in terms of the Minkowski convention $\gamma^3$ to match the conventions of other quantities used in this appendix. Beginning with Fig.~\ref{fig:diags2}(a), we first evaluate the isospin projected result with the spin up $N \pi$ state
\begin{align}
\begin{split}
&\langle N \pi, +,\mathrm{out} \vert A^3_3(0) \vert N \rangle \\
 & \hspace{15pt} \supset  i \frac{2}{\sqrt{3}} \overline u_+(\textbf 0) e^{- i \eta \hat p \cdot \vec K }   \left(  \frac{i}{2f_\pi}\gamma^3 \right )    u_+(\textbf 0) 
 \end{split}  \\ 
 &   \hspace{15pt}  =    \frac{  \sinh(\eta/2)  \hat p_i  }{\sqrt{3} f_\pi  }  \ \overline u_+(\textbf 0) \gamma_0 \gamma^i   \gamma^3   u_+(\textbf 0)   \\ 
   &    \hspace{15pt}  =-     \frac{2}{\sqrt{3}}  \frac{   \sqrt{m_N} (\omega_{N} - m_N)^{1/2} }{ \sqrt{6}f_\pi  }      \sqrt{4 \pi} Y_{10}(\hat p) \,. % \\ 
\end{align}

The contribution from the spin down state has a similar form
\begin{multline}
\langle N \pi, -,\mathrm{out} \vert A^3_3(0) \vert N \rangle \\
 \supset  2 \sqrt{ \frac{2}{3} } \frac{ \sqrt{m_N}(\omega_{N} - m_N)^{1/2} }{ \sqrt{6} f_\pi  }  \sqrt{4 \pi} Y_{11}(\hat p) \,,
\end{multline}
and the final result for the Fig.~\ref{fig:diags2}(a) contribution to the $J=1/2$ matrix element is
\begin{equation}
\langle N \pi, \mathrm{out} \vert A^3_3(0) \vert N \rangle \supset  2   \frac{ \sqrt{m_N}(\omega_{N} - m_N)^{1/2} }{ \sqrt{6} f_\pi  }   \,.
\end{equation}

Next evaluating Fig.~\ref{fig:diags2}(b) we find
\begin{align}
\begin{split}
 & \langle N \pi, + , \mathrm{out} \vert A^3_3(0) \vert N \rangle \supset   \sqrt{3} \overline u_+(\textbf 0) e^{- i \eta \hat p \cdot \vec K }  \\ &  \times   \left( i \frac{g_A}{2f_\pi} i \slashed k \gamma_5 \right )  i \frac{(\omega_N + \omega_\pi) \gamma^0 + m_N}{(\omega_N + \omega_\pi)^2 - m_N^2 } \\ & \times \left(- i \frac{g_A}{2} \gamma^3 \gamma_5 \right )                  u_+(\textbf 0) 
 \end{split} \\
 & =  \frac{\sqrt{2}  g_A \bar g_A  \sqrt{m_N}  (\omega _N-m_N)^{1/2} }{4f_\pi }   \sqrt{4 \pi} Y_{10}(\hat p) 
\end{align}
for the spin-up matrix element,
\begin{multline}
 \langle N \pi, - , \mathrm{out} \vert A^3_3(0) \vert N \rangle \supset \\  - \frac{  g_A \bar g_A  \sqrt{m_N}  (\omega _N-m_N)^{1/2} }{2f_\pi }   \sqrt{4 \pi} Y_{11}(\hat p) 
\end{multline}
for the spin-down matrix element, and
\begin{equation}
 \langle N \pi, \mathrm{out} \vert A^3_3(0) \vert N \rangle \supset  -  \frac{3  \sqrt{m_N}(\omega_{N} - m_N)^{1/2}  g_A \bar g_A}{  2 \sqrt{6} f_\pi  }      
\end{equation}
for the $J=1/2$ matrix element.

We conclude with Fig.~\ref{fig:diags2}(c), jumping straight to the projected, $J=1/2$ result 
\begin{multline}
 \langle N \pi, \mathrm{out} \vert A^3_3(0) \vert N \rangle \supset    \frac{  \sqrt{m_N}(\omega_{N} - m_N)^{1/2}  }{  6 \sqrt{6} f_\pi  }  g_A   \\ \times \left[    \bar g_A  + g_A \frac{4   M_\pi^2}{  2 \omega_{\pi} m_N-  M_\pi^2 }  \right ] \,.
\end{multline}
Combining all results, we conclude
\begin{multline}
 \langle N \pi, \mathrm{out} \vert A^3_3(0) \vert N \rangle =   \frac{ \sqrt{m_N}(\omega_{N} - m_N)^{1/2} }{ 2 \sqrt{6} f_\pi  }     \\
  \times  \left[4-\frac{8}{3} g_A \left(\overline g_A-\frac{g_A M_\pi^2}{  4 \omega_{\pi} m_N-2 M_\pi^2 }\right)\right] \,.
  \label{eq:axialME}
\end{multline}

Substituting Eq.~(\ref{eq:axialME}) into Eq.~(\ref{eq:bLL}) then gives
\begin{multline}
b_n =  \frac{\nu_n}{16     \omega_{\pi } f_\pi^2  L^3  }    \left (1 - \frac{m_N}{\omega_N} \right)  (1 - \bar g_A)  \\ \times   
    \left[4-\frac{8}{3} g_A \left(\overline g_A-\frac{g_A M_\pi^2}{  4 \omega_{\pi} m_N-2 M_\pi^2 }\right)\right]  \,,
\end{multline}
in perfect agreement with Ref.~\cite{BarGA}.

\bibliography{MyLibrary2} %%% ref.bib file
\end{document}